\documentclass[12pt]{iopart}
\usepackage{graphicx}
\usepackage[utf8]{inputenc}
\usepackage[T1]{fontenc}

\expandafter\let\csname equation*\endcsname\relax
\expandafter\let\csname endequation*\endcsname\relax
\usepackage{amsmath,amsfonts}
\usepackage{hyperref}
\usepackage[capitalize]{cleveref}
\usepackage{xcolor}
\usepackage{soul}
\usepackage{subfigure}
\usepackage[normalem]{ulem}

\begin{document}

\setstcolor{red}

\title[Single-Stage Stellarator Optimization with Fixed Boundary Equilibria]{Single-Stage Stellarator Optimization: Combining Coils with Fixed Boundary Equilibria}

\author{R. Jorge}
\address{Instituto de Plasmas e Fusão Nuclear, Instituto Superior Técnico, Universidade de Lisboa, 1049-001 Lisboa, Portugal}
\address{Departamento de Física, Instituto Superior Técnico, Universidade de Lisboa, 1049-001 Lisboa, Portugal}
\ead{rogerio.jorge@tecnico.ulisboa.pt}

\author{A. Goodman}
\address{Max-Planck-Institut für Plasmaphysik, D-17491 Greifswald, Germany}

\author{M. Landreman}
\address{Institute for Research in Electronics and Applied Physics, University of Maryland, College Park, MD 20742, USA}

\author{J. Rodrigues}
\address{Departamento de Física, Instituto Superior Técnico, Universidade de Lisboa, 1049-001 Lisboa, Portugal}

\author{F. Wechsung}
\address{Courant Institute of Mathematical Sciences, New York University, New York, NY, 10012, USA}

\begin{abstract}
{We introduce a novel approach for the simultaneous optimization of plasma physics and coil engineering objectives using fixed-boundary equilibria that is computationally efficient and applicable to a broad range of vacuum and finite plasma pressure scenarios. Our approach treats the plasma boundary and coil shapes as independently optimized variables, penalizing the mismatch between the two using a quadratic flux term in the objective function. Four use cases are presented to demonstrate the effectiveness of the approach, including simple and complex stellarator geometries. As shown here, this method outperforms previous 2-stage approaches, achieving smaller plasma objective function values when coils are taken into account.}
\end{abstract}

\section{Introduction}
\label{sec:intro}

Developing a practical and economically viable fusion device has proven to be a formidable challenge.
Several factors, including plasma confinement and heating, plasma stability, and the efficiency of energy conversion determine the performance of nuclear fusion devices.
In order to achieve the conditions necessary for nuclear fusion to occur, the plasma must be confined by a magnetic field at high temperatures and densities for long periods of time.
The magnetic field configuration plays a critical role in achieving these conditions, and the design of the coils that generate the magnetic field is a crucial aspect of this challenge.
Over the years, various device designs with the goal of developing a viable fusion power plant have been studied.
One of the most promising devices is the stellarator, a type of magnetic confinement device that uses a complex system of coils to create a magnetic field and has the ability to operate in a steady-state mode without a disruptive limit \cite{Helander2014,Imbert-Gerard2019}.
Compared to tokamaks, stellarators have simpler plasma control, require less injected power to sustain the plasma since the current drive is unnecessary, and can have a very flexible plasma shape.
Although this flexibility is theoretically desirable, it comes at the cost of much-increased complexity in the coils, as the plasma usually needs to be shaped in a complex way to achieve good performance, therefore increasing the cost of the physical devices and hindering their economic viability. 
Hence, to make the construction of these devices more economically viable, it is important to explore possible methods for optimizing coil designs while maintaining the performance of the underlying magnetic field equilibria.

In this work, we present a new method for performing a combined plasma coil optimization algorithm using fixed boundary equilibria.
This method uses a combined plasma-coil optimization in a single-stage approach that takes into account both physics goals and engineering constraints simultaneously.
As we show here, this method enables us to achieve smaller values of the plasma objective (in the present case, quasisymmetric or quasi-isodynamic objectives) with coils than the standard two-stage approach.
This overcomes the challenges of previous methods based on free boundary equilibria and can be applied to arbitrary stellarator equilibria.
Up to now, the standard technique used in the design process of the stellarator is a two-stage approach.
The first step of this approach focuses on determining the desired properties of the target magnetic field equilibrium, such as its aspect ratio, quasisymmetry, MHD stability, and the properties of its magnetic islands.
The second stage consists in finding a set of coils that are able to recreate that target field.
The two-stage approach is commonly used today since it is an efficient technique that commonly leads to the toroidal surfaces foliating a large fraction of the plasma volume and has led to the design of many successful stellarator experiments.
As there may be many different sets of coils that can produce the same target magnetic field, the second stage is an ill-posed problem.
This can make it challenging to determine the optimal coil configuration for a given stellarator, because a slight modification in the plasma boundary may demand a significant adjustment to the coil geometry.
Therefore, since the target magnetic field is fixed, the set of coils that are found can present an unrealistic challenge to the fabricator, as a result of the very high complexity.
This would mean that the process would have to restart, requiring another target field, again without any certainty that it would lead to a realistic coil design.

These challenges and difficulties may be resolved if a single-stage optimization approach is considered, in which the target magnetic field and its accompanying coils are varied at the same time.
This way, the coil complexity can be balanced with the plasma performance, with the goal of achieving good confinement and stability without sacrificing engineering feasibility.
In this case, at each iteration step, the plasma equilibrium, the magnetic field from the coils, and the current contribution from the plasma are evaluated simultaneously.
This joint optimization approach makes it possible to achieve the desired balance between complexity and performance in an efficient way.

Previous combined optimization methods have either relied directly or indirectly on free boundary equilibrium calculations \cite{Strickler2002,Hudson2002,Strickler2017,Reiman2017,Hudson2018}, which often demand many iterations between an equilibrium solution, or these can only be used with vacuum configurations \cite{Yamaguchi2019,Yu2021,Yu2022,Giuliani2022b,Giuliani2022c}.
In Ref. \cite{Henneberg2021} several combined optimization approaches were discussed, including the possibility of using fixed boundary equilibria and the corresponding penalty functionals. While the present work is similar in nature to Section 5.1 of Ref. \cite{Henneberg2021} for a general objective function, here, we consider the degrees of freedom for the optimization to be both the plasma boundary and coil degrees of freedom and define the objective function to be the linear combination of both the stage 1 and stage 2 objective functions in order to achieve our goal of good confinement and simpler coils.
Those reasons support the choice of this particular single-stage approach.
As a concrete example, we take the latest optimization stage of the NCSX device where the coil shapes were directly included in the optimization of the plasma shape \cite{Reiman2017}.
In this case, a combined plasma-coil algorithm based on a free-boundary equilibrium was used after a two-stage optimization process that had already identified a good candidate for the design point, thus providing an initial guess for the local minimization that includes both plasma and coil models.
The degrees of freedom used were the parameters describing the coil shapes and the coil currents and the target included physics parameters of the reference plasma and the geometric properties necessary for engineering coil design.
As stated in Ref. \cite{Strickl2003}, it was not until a combined plasma-coil optimization was performed that a family of consistent solutions (termed M45) that met engineering feasibility requirements and adequately reconstructed the plasma properties of the initial LI383 equilibrium was found.
Another example where an \textit{a posteriori} combined plasma-coil optimization led to major improvements in loss fraction and effective helical ripple with respect to an already found solution from a two-stage approach is described in Ref. \cite{Giuliani2022b}, although this method is only applicable to quasisymmetric magnetic fields in vacuum.
Such studies show the importance of combined plasma-coil optimization algorithms in the design of stellarator experiments.

Our new method introduces a streamlined approach for combined plasma-coil optimization in stellarator design, which enables simultaneous optimization of both the physics and coil engineering objectives in a single stage.
At the core of our new method is the principle of including both the plasma boundary shape and coil shapes in the optimization parameter space as degrees of freedom.
At the same time, we introduce a quadratic flux \cite{Dewar1994} term in the objective function to ensure consistency between the two, similar to what is done in Ref. \cite{Giuliani2022a} using the near-axis expansion approach{, i.e., obtaining the magnetic field equilibrium using an expansion at successive orders in the distance from the axis \cite{Garren1991,Landreman2018,Jorge2020}}.
{The quadratic flux term, defined in \cref{eq:quadraticflux}, is the surface integral of the normal component of the magnetic field produced from the coils which is zero for the ideal equilibrium case.}
We combine finite difference derivatives of the MHD equilibrium with analytic derivatives of the coils, resulting in a reduced number of finite difference steps.
Additionally, our approach can be applied to equilibrium codes that do not yet have free boundary functionality, such as GVEC \cite{Maurer2020}, making it adaptable to a broad range of vacuum and finite plasma pressure stellarator equilibria.
We note that, in this method, only one surface evaluation of the magnetic field from coils is required per optimization iteration, significantly reducing the computational time as opposed to methods that volumetrically evaluate the magnetic field such as the free-boundary version of the VMEC code \cite{Hirshman1983} where an mgrid file from the MAKEGRID code is needed.

The paper is organized as follows.
In \cref{sec:physics}, we provide an overview of the method and describe the numerical implementation and the codes used.
In \cref{sec:convergence}, we verify our method by performing convergence studies on the optimization objective function and its gradients.
The results of our approach when applied to quasisymmetric and quasi-isodynamic configurations are shown in \cref{sec:results}.
The conclusions follow.

\section{Optimization Method}
\label{sec:physics}

The magnetic field equilibrium is obtained using VMEC (Variational Moments Equilibrium Code) \cite{Hirshman1983}, which solves the static ideal magnetohydrodynamics (MHD) system of equations
\begin{equation}
    \mathbf J \times \mathbf B = \nabla P,
\label{eq:MHD}
\end{equation}
where $\mu_0 \mathbf J = \nabla \times B$ is the plasma current density, $\mathbf B$ the equilibrium magnetic field satisfying $\nabla \cdot \mathbf B = 0$ and $P$ the plasma pressure.
The ideal MHD model is valid on a low-frequency and long-wavelength regimes, where typical frequencies $\omega$ are larger than $\omega_p$, the plasma frequency, and larger than the electron and ion gyrofrequencies $\Omega_{e,i}$, and where typical length scales $L$ are longer than the Debye length $\lambda_D$ and the electron and ion gyroradii $\rho_{e,i}$ \cite{Freidberg2014}.
Furthermore, it is assumed that collisions are frequent enough for the electron and ion distribution functions to thermalize.
VMEC assumes a toroidal equilibrium with nested surfaces of constant toroidal magnetic flux, also called flux surfaces, and uses the steepest descent method to find a minimum in the potential energy $W$ resulting from an integral formulation of \cref{eq:MHD}, namely
\begin{equation}
    W=\int\left(\frac{|B|^2}{2 \mu_0}+\frac{p}{\Gamma-1}\right)dV,
\end{equation}
where $V$ is the integration volume and $\Gamma=5/3$ is the ratio of specific heats.

We run VMEC in fixed boundary mode.
In this case, the outermost surface $S=[R(\vartheta,\phi)\cos(\phi), R(\vartheta,\phi)\sin(\phi), Z(\vartheta,\phi)]$, also called last closed flux surface, is fixed and used as a boundary condition. The boundary surface $S$ is specified by its Fourier amplitudes $\{\mathrm{RBC}_{m,n},\mathrm{ZBS}_{m,n}\}$ in cylindrical coordinates
\begin{equation}
    R(\vartheta,\phi)=\sum_{m=0}^{M_{\mathrm{pol}}}\sum_{n=-N_{\mathrm{tor}}}^{N_{\mathrm{tor}}}\mathrm{RBC}_{m,n}\cos(m\vartheta-n_{\text{fp}}n\phi),
\end{equation}
and
\begin{equation}
    Z(\vartheta,\phi)=\sum_{m=0}^{M_{\mathrm{pol}}}\sum_{n=-N_{\mathrm{tor}}}^{N_{\mathrm{tor}}}\mathrm{ZBS}_{m,n}\sin(m\vartheta-n_{\text{fp}}n\phi),
\end{equation}
where $\phi$ is the standard cylindrical angle, $\vartheta$ is a poloidal angle, and $n_{\text{fp}}$ is the number of toroidal field periods of the magnetic field equilibrium.
Only $\cos$ and $\sin$ modes are used to specify $R$ and $Z$ respectively, so as to enforce stellarator-symmetry throughout this work.
At each magnetic surface, the toroidal magnetic flux $2\pi \psi$ is constant.
We denote by $\psi_b$ the value of $\psi$ at the plasma boundary and $s=\psi/\psi_b$ the normalized toroidal flux.
The degrees of freedom for the surface shapes are then
\begin{equation}
    \mathbf x_{\text{surface}}=\left[\text{RBC}_{m,n},\text{ZBS}_{m,n}\right].
\end{equation}
We mention two important properties of the surface $S$ that we use throughout this work, namely its normal vector $\mathbf n(\vartheta,\phi)$ and its aspect ratio $A$ defined as
\begin{equation}
    A = \frac{\texttt{Rmajor\_p}}{\texttt{Aminor\_p}}=\frac{V}{2\pi^2\texttt{Aminor\_p}^3}=\frac{V}{2\sqrt{\pi}\overline S^{3/2}},
\end{equation}
where $\overline S = (2\pi)^{-1}\int_0^{2\pi}d\phi S(\phi)$ is the toroidal average of the area $S(\phi)$ of the outer surface's cross section in the $R-Z$ plane and $V$ is the volume of the outer surface \cite{Landreman2019b}.

To illustrate how a physics property can be targeted while simultaneously optimizing coil shapes, we choose precise quasisymmetric \cite{Nuhrenberg1988} and quasi-isodynamic \cite{Gori1994} magnetic fields as targets for the equilibrium magnetic field.
Quasisymmetry is one of the ways to achieve the good confinement properties of a tokamak with the stability and steady-state capability of a stellarator.
In particular, if the modulus of the magnetic field vector $|\mathbf B|=B$ has the following symmetry
\begin{equation}
    B=B(\psi, M \theta - N \varphi),
\label{eq:Bboozer}
\end{equation}
where $(\theta,\varphi)$ are Boozer coordinates \cite{Boozer1981}.
This results in trajectories of the guiding center of charged particles that behave exactly as if they were in a truly symmetric magnetic field vector $\mathbf B$.
Note that $\mathbf B$ is not required to have any particular symmetry and this symmetry is only dependent on the surface degrees of freedom $\mathbf x_{\text{surface}}$ if a fixed boundary approach is employed.
To achieve quasisymmetry, we follow the approach in \cite{Landreman2022} and rewrite \cref{eq:Bboozer} in an equivalent form, namely \cite{Jorge2020a}
\begin{equation}
    \mathbf B \times \nabla \psi \cdot \nabla B = F(\psi) \mathbf B \cdot \nabla B,
\end{equation}
where $F(\psi)=(M G+ N I)/(\iota M - N)$ if $\mathbf B$ is expressed in Boozer coordinates
and seek the minimization of the objective function $f_{\text{QS}}$
\begin{align}
    f_{\text{QS}} = &\sum_{s_j}\left<\left(\frac{1}{B^3}\left[(N-\iota M)\mathbf B \times \nabla B \cdot \nabla \psi-(MG+NI)\mathbf B \cdot \nabla B\right]\right)^2\right>,
\label{eq:fqs}
\end{align}
where $G(\psi)$ is $\mu_0/(2\pi)$ times the poloidal current outside the surface, $I(\psi)$ is $\mu_0/(2\pi)$ times the toroidal current inside the surface, $\iota$ is the rotational transform and $\left<\dots\right>$ is a flux surface average.
The sum is over a set of flux surfaces $s_j=\psi_j/\psi_b$ where $\psi_b$ is the toroidal flux at the boundary and a uniform grid $0, 0.1, \dots, 1$ is used.
The quantities $\mathbf B \times \nabla B \cdot \nabla \psi$, $\mathbf B \cdot \nabla B$, $B$, $G$ and $I$ are computed using VMEC, while we set
\begin{equation}
    (M,N) = 
  \begin{cases}
    (1,0), & \text{for quasi-axisymmetry},\\
    (1,-1), & \text{for quasi-helical symmetry},
  \end{cases}
\end{equation}
which are the two allowed flavors of quasisymmetry close to the magnetic axis \cite{Plunk2019}.
{We use this particular form of quasisymmetry, \cref{eq:fqs},  primarily due to its accessibility within the SIMSOPT framework and its demonstrated smoothness and convergence properties. Nevertheless, further exploration of different quasisymmetry forms (e.g., \cite{Dudt2023}), might contribute to the refinement of the optimization method applied in the current work}
Quasi-isodynamic magnetic fields are fields that are omnigeneous with poloidally closed contours of the magnetic field strength $B$.
To achieve a quasi-isodynamic magnetic field, we follow the approach in \cite{Goodman2022} and
\begin{equation}
    f_\text{QI}=\frac{n_\text{fp}}{4 \pi^2}\sum_{s_j}\int_0^{2\pi}d\alpha\int_0^{2\pi/n_{\text{fp}}}d\varphi\left(\frac{B-B_{QI}}{B_{\text{max}}-B_{\text{min}}}\right)^2,
\end{equation}
where $B_{QI}$ is the target magnetic field, $\alpha$ is a fieldline label that in Boozer coordinates can be written as $\alpha=\vartheta-\iota \varphi$, and $B_\text{max}$ ($B_\text{min}$) is the maximum (minimum) magnetic field strength on a flux surface.
As QI-optimized magnetic fields often result in elongated flux surfaces and large differences between the maximum and minimum magnetic field strengths on a given flux surface, we also add to the objective function a penalty on the effective elongation $\epsilon$ defined in \cite{Goodman2022} and the mirror ratio $\Delta$ defined as
\begin{equation}
    \Delta = \frac{B_\text{max}-B_\text{min}}{B_\text{max}+B_\text{min}}.
\end{equation}

The objective function for the equilibrium magnetic field $J_1$, when targeting quasisymmetry, is then written as 
\begin{align}
    J_1 = &f_{\text{QS}}+(A-A_{\text{target}})^2 + c(\iota).
\label{eq:J1_QI}
\end{align}
The target aspect ratio $A_{\text{target}}$ is set to $6$ for quasi-axisymmetric stellarators and to $8$ for quasi-helically symmetric stellarators.
The function $c(\iota)$ places a constraint on the rotational transform profile in order to restrict quasi-axisymmetric configurations to not become axisymmetric.
We therefore take $c=(\text{mean}(\iota)-\iota_0)^2$ for quasi-axisymmetry and $c=0$ for quasi-helical symmetry optimization.
The objective function for the equilibrium magnetic field $J_1$, when targeting quasi-isodynamic magnetic fields, is then written as 
\begin{align}
    J_1 = &f_{\text{QI}}+(A-A_{\text{target}})^2 + \text{max}(0,\Delta-\Delta_*)^2+\text{max}(0,\epsilon-\epsilon_*)^2.
\label{eq:J1}
\end{align}
The maximum allowed mirror ratio $\Delta_*$ is chosen to be $\Delta_*=0.21$ and the maximum allowed elongation $\epsilon_*=6.0$.

We then place a set of $N_C$ electromagnetic coils surrounding half of a field period of the plasma boundary since, due to stellarator-symmetry and $n_{\text{fp}}$ rotational symmetry, the remaining coils can be found by rotation and reflection transformations of the independent $N_C$ coils.
The total number of coils is then given by $2n_{\text{fp}}N_C$.
While we only use modular toroidal field coils, other types of coils can be used in the framework proposed here such as saddle coils and poloidal field coils.
The coils are represented as current-carrying filaments, i.e., the non-zero thickness of the coils is neglected and they are modeled as curves in space, a common assumption in coil design \cite{Zhu2017a}.
The magnetic field from the coils is evaluated using the Biot-Savart law 
\begin{equation}
    \mathbf B_{\text{ext}}(\overline{\mathbf x})=\frac{\mu_0}{4\pi}\sum_{i=1}^{2n_{\text{fp}}N_C} I_i \int_{\mathbf \Gamma_i}\frac{d \mathbf l_i\times \mathbf r}{r^3},
\end{equation}
where $I_i$ is the current in the $i$th coil $\mathbf \Gamma_i$, $d \mathbf l_i = \mathbf x'_i d\theta$ is the differential line element, $\theta$ is an angle-like coordinate that parametrizes the coil curve $\mathbf \Gamma_i$ and $\mathbf r=\overline{\mathbf x} - \mathbf x_i$ is the displacement vector between the evaluated point on the surface and the differential element.
Each coil $i$ is modeled as a periodic function
\begin{equation}
    \mathbf \Gamma^{(i)}=[\Gamma_1^{(i)},\Gamma_2^{(i)},\Gamma_3^{(i)}]:[0,2\pi)\rightarrow \mathbb{R}^3,
\end{equation}
where
\begin{equation}
\Gamma_j^{(i)}=c_{j,0}^{(i)}+\sum_{l=1}^{N_F}\left[c_{j,l}^{(i)}\cos(l \theta)+s_{j,l}^{(i)}\sin(l \theta)\right],
\end{equation}
yielding a total of $3(2N_F +1)$ degrees of freedom per coil.
The degrees of freedom for the coil shapes are then
\begin{equation}
    \mathbf x_{\text{coils}}=[c_{j,l}^{(i)},s_{j,l}^{(i)}, I_i].
\end{equation}
To prevent the minimization of the quadratic flux by all coil currents going to zero, we take one of the coil currents $I_i$ out of the parameter space.

To find suitable coils, we follow the approach of the FOCUS coil design tool \cite{Zhu2017a} and vary $\mathbf x_{\text{coils}}$ in order to minimize the field error, i.e., the magnitude of the normal component $\mathbf n$ of the magnetic field induced by the coils on the boundary $S$.
This is in contrast to the current carrying surface approach where coils are restricted to lie on a particular surface (called the coil winding surface) which is employed in the NESCOIL \cite{Merkel1987}, REGCOIL \cite{Landreman2017}, ONSET \cite{Drevlak1999}, COILOPT \cite{Strickler2002}, and COILOPT++ \cite{Gates2017} codes.
The FOCUS approach implemented in SIMSOPT has been shown to yield exceptionally low field errors and allow for simple coil shapes (see Refs. \cite{Wechsung2022,Giuliani2022b}) and is therefore the one used here.
We target the minimization of the field error by defining as a cost function the quadratic flux quantity $f_{\text{QF}}$, given by
\begin{equation}
    f_{\text{QF}}=\int_S \left(\frac{\mathbf B_{\text{ext}} \cdot \mathbf n}{|\mathbf B_{\text{ext}}|}\right)^2 dS,
\label{eq:quadraticflux}
\end{equation}
where $\mathbf B_{\text{ext}}=\mathbf B_{\text{ext}}(\mathbf x_{\text{coils}})$, $\mathbf n = \mathbf n(S)$ and $S=S(\mathbf x_{\text{surface}})$.
If the field induced by the coils $\mathbf B_{\text{ext}}$ coioncides with the target equilibrium field $\mathbf B$, then $f_{\text{QF}}=0$.
However, minimizing $\mathbf x_{\text{coils}} \rightarrow f_{\text{QF}}(\mathbf B_{\text{ext}}(\mathbf x_{\text{coils}}))$ is an ill-posed problem \cite{Landreman2017,Zhu2017a}.
For this reason, we restrict the allowable coil shapes by penalizing coil complexity using the method introduced in Ref. \cite{Wechsung2022}.
In this way, we obtain a design where it is plausible that the coils can be built.
Therefore, we consider the following regularization terms
\begin{align}
    g_L&=\varphi\left(\sum_{i=1}^{N_c}L_i-L_{\text{max}}\right),\\
    g_{\kappa,\text{max}}&=\sum_{i=1}^{N_c}\int_0^{2\pi}\mathrm{M}(\kappa_i-\kappa_{\text{max}})|\Gamma^{'(i)}|d\theta/L_i,\\
    g_{\kappa,\text{msc}}&=\sum_{i=1}^{N_c}\varphi\left(\int_0^{2\pi}\kappa_i^2|\Gamma^{'(i)}|d\theta/L_i-\kappa_{\text{msc}}\right),\\
    g_d&=\sum_{i=1}^{N_C}\sum_{j=1}^{i-1}\int_0^{2\pi}\int_0^{2\pi}\mathrm{M}(d_{\text{min}}-|\Gamma^{(i)}(\theta)-\Gamma^{(j)}(\theta')|)|\Gamma^{'(i)}(\theta)\Gamma^{'(j)}(\theta')|d \theta d \theta',\\
    g_\ell&=\text{Var}(\{\ell_j^{(i)}\}^{2N_F-1}_{j=0}).
\end{align}
where $L_i$ and $\kappa_i$ are the length and curvature of the coil $i$, respectively, $\mathrm{M}(t)=\text{max}(t,0)^2$ and $\ell_j^{(i)}$ is the variance of the coil arclength.
These allow us to restrict the total length of the independent coils to be $L_{\text{max}}$, the individual coil curvature and mean squared curvature to $\kappa_{\text{max}}$ and $\kappa_{\text{msc}}$, respectively, and the minimum distance between coils to $d_{\text{min}}$.
The regularization term $g_\ell$ is added to avoid poor conditioning of the optimization problem due to  non-uniqueness of the curve parameter by enforcing uniform arclength along the curve \cite{Wechsung2022}.
In essence, such regularization terms intend to restrict the space of allowed coil shapes to satisfy the following set of conditions
\begin{subequations} \label{eq:opt_conds}
\begin{align}
    \sum_{i=1}^{N_c}L_{c}^{(i)}&\le L_{\text{max}},\\
    \kappa_i &\le \kappa_{\text{max}},~i=1,\ldots,N_c,\\
    \frac{1}{L_c^{(i)}}\int_{\gamma^{(i)}}\kappa_i^2 dl &\le \kappa_{\text{msc}},~i=1,\ldots,N_c,\\
    \| \Gamma^{(i)}-\Gamma^{(j)} \| &\geq d_{\min} ~ \text{ for } i \neq j,\label{eq:optprob:8}.
\end{align}
\end{subequations}
The objective function for the coils $J_2$ is then written as 
\begin{align}
    J_2 &= f_{\text{QF}}+\omega_L g_L+\omega_{\kappa,\text{max}} g_{\kappa,\text{max}}+\omega_{\kappa,\text{msc}} g_{\kappa,\text{msc}}+\omega_d g_d + \omega_\ell g_\ell,
\label{eq:J2}
\end{align}
where $\omega_{g_i}$ are the scalar weights associated with the regularization terms $g_i$.

The optimization problem and total objective function $J$ can then be formulated as
\begin{subequations} \label{eq:opt}
\begin{align}
    \min_{\mathbf x_{\text{coils}},\mathbf x_{\text{surface}}} &J(\mathbf x_{\text{coils}},\mathbf x_{\text{surface}})=J_1+\omega_{\text{coils}}J_2, \label{eq:J}\\
    \text{subject to } &~\psi=\psi_0,~R_{\text{major}}=R_0,
\end{align}
\end{subequations}
with $\omega_{\text{coils}}$ a scalar weight associated with $J_2$.
The degrees of freedom varied during the optimization are $\mathbf x=(\mathbf x_{\text{surface}}, \mathbf x_{\text{coils}})$.
The objective function, its gradients, as well as its CPU parallelization, are carried out using the SIMSOPT code \cite{Landreman2021b}.
The constraint $\psi=\psi_0$ is handled by the use of VMEC in fixed-boundary mode, while the constraint $R_{\text{major}}=R_0$ is handled by removing the term $\mathrm{RBC}_{0,0}$ from the parameter space and setting it equal to one.

In the numerical results section, we minimize $J$ by employing the Broyden–Fletcher–Goldfarb–Shanno (BFGS) quasi-Newton algorithm \cite{Nocedal2006}.
This method stores the Jacobian of the objective function computed at previous points and uses the BFGS formula to approximate the Hessian using a generalized secant method.
As noted in \cite{Wechsung2022}, this is important for convergence as it helps to overcome the ill-posedness of the coil optimization problem.
The Jacobian $dJ/d\mathbf x=dJ_1/d\mathbf x+dJ_2/d\mathbf x$ is computed using a mix of numerical and analytical derivatives.
In particular, the Jacobian $dJ_1/d\mathbf x_{\text{surface}}$ is computed using forward finite differences, $dJ_1/d\mathbf x_{\text{coils}}=0$, and both $dJ_2/d\mathbf x_{\text{coils}}$ and $dJ_2/d\mathbf x_{\text{surface}}$ are computed analytically.
The use of analytical derivatives allows significant efficiency in the optimization process as the number of degrees of freedom $\mathbf x_{\text{coils}}$ is typically significantly larger than $\mathbf x_{\text{surface}}$.
As an example, while a surface with $M_{\text{pol}}=3$ poloidal and $N_{\text{tor}}=3$ toroidal modes has 49 degrees of freedom (length of $\mathbf x_{\text{surface}}$), a system with $N_C=4$ independent coils and $N_F=16$ Fourier modes per coil has a total of 396 degrees of freedom (length of $\mathbf x_{\text{coils}}$).
It is worth mentioning that the method introduced here is applicable (and could entail significant advantages) even in the case where analytic derivatives of the equilibrium are available, using codes such as DESC \cite{Dudt2020}.
{
Although the number of degrees of freedom in coils is generally larger than that in surfaces, it is essential to note that specific cases may vary. Specifically, the four-coil stellarator outlined in this work has a smaller number of degrees of freedom than the surface shape.
}

\section{Gradient Verification}
\label{sec:convergence}

We now turn to the numerical validation of the approach implemented here which will focus on the computation of the gradients of $J$.
This is done for several reasons with the major one being the fact that the main modifications of the SIMSOPT code were performed in the functions directly or indirectly related to the computation of the gradients.
Furthermore, the gradients of $J$ are a crucial element in the optimization algorithm, making its accuracy extremely relevant.
Finally, gradients are relevant to the present work as a significant efficiency gain in the approach used here, apart from the use of fixed boundary equilibria, is the use of analytical derivatives for the computation of the derivatives of $J_2$ and the parallelization of the finite difference gradients of $J_1$.

We start by performing a Taylor test to verify the accuracy of the analytical derivatives on the computation of the objective function $J_2$ with respect to the degrees of freedom $\mathbf x$.
For this purpose, we estimate $dJ_2/d\mathbf x\simeq \Delta J_2/\Delta x$ using a finite-difference approach
\begin{equation}
    \frac{\Delta J_2}{\Delta x} = \frac{J_2(\mathbf x + \mathbf h \Delta x)-J_2(\mathbf x - \mathbf h \Delta x)}{2 \Delta x}
\label{eq:estimate_dJ2}
\end{equation}
where $\mathbf h$ is a randomized array with the same size as $\mathbf x$ and with values between $0$ and $1$.
We then compare it to the analytical estimate $J_2'(\mathbf x)$ provided by the code for several values of $\delta x$ and assess its convergence.

\begin{figure}
    \centering
    \includegraphics[width=.6\textwidth]{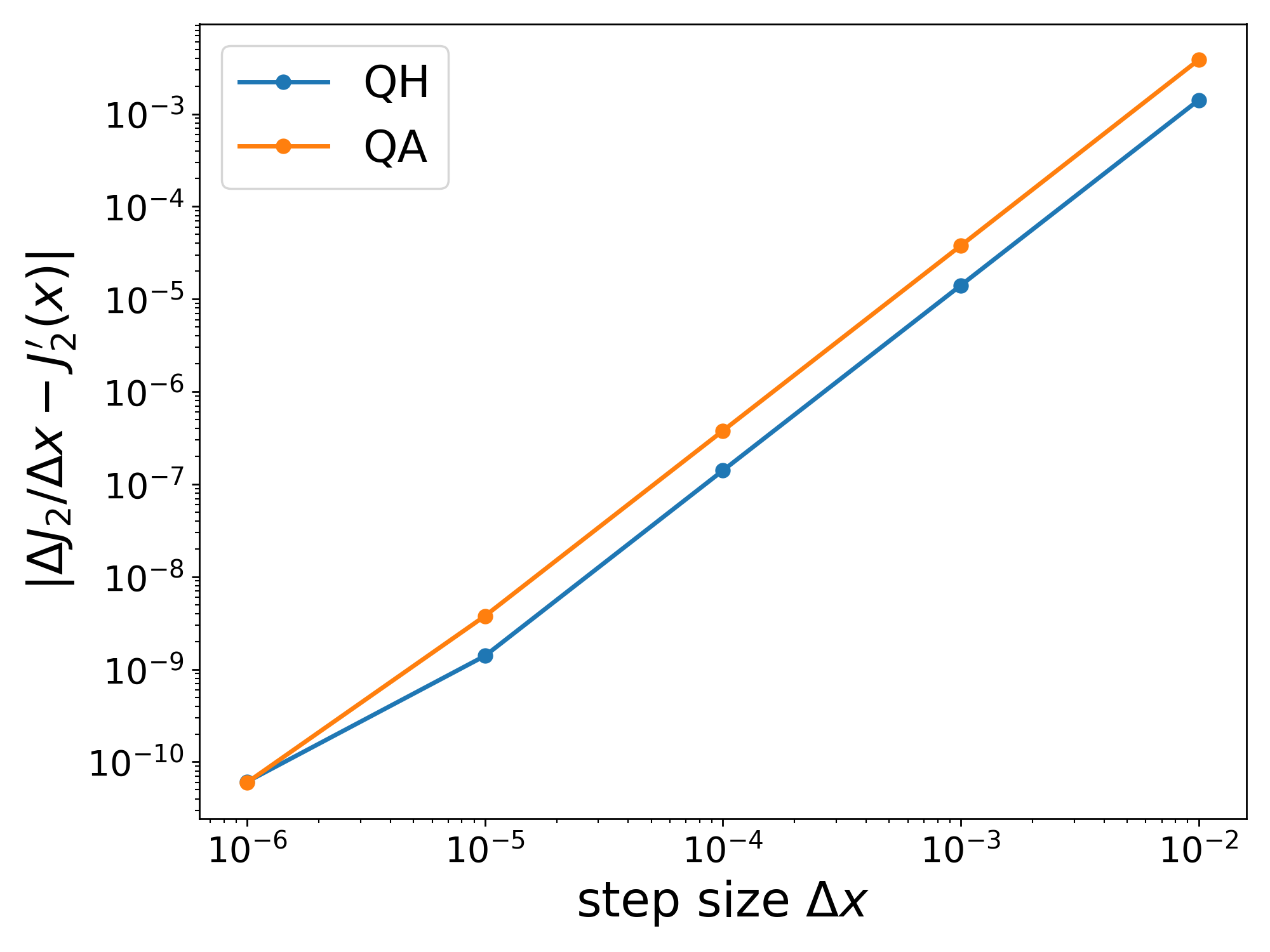}
    \caption{Second order convergence observed for the difference between the analytical estimate $J_2'(\mathbf x)$ and the centered finite difference estimate $\Delta J_2/\Delta x$ employed in \cref{eq:estimate_dJ2}. The boundary surfaces are the QA and QH configurations of Ref. \cite{Landreman2022} and the coils consist of three curves per half field period with a major radius of 1 m and a minor radius of 0.5 m.}
    \label{fig:estimate_dJ2}
\end{figure}

This is shown in \cref{fig:estimate_dJ2} where second-order convergence $|\Delta J_2/\Delta x-J_2'(\mathbf x)| \sim |\Delta x|^2$ is obtained as expected from the use of a centered finite difference scheme in \cref{eq:estimate_dJ2}.
The coils employed in this study are a set of three circular coils per half-field period with a major radius of 1 m and a minor radius of 0.5 m.
The plasma boundaries used are the ones present in the precise quasi-axisymmetry (QA) and precise quasi-helically symmetric (QH) configurations of Ref. \cite{Landreman2022}.

\begin{figure}
    \centering
    \includegraphics[width=.49\textwidth]{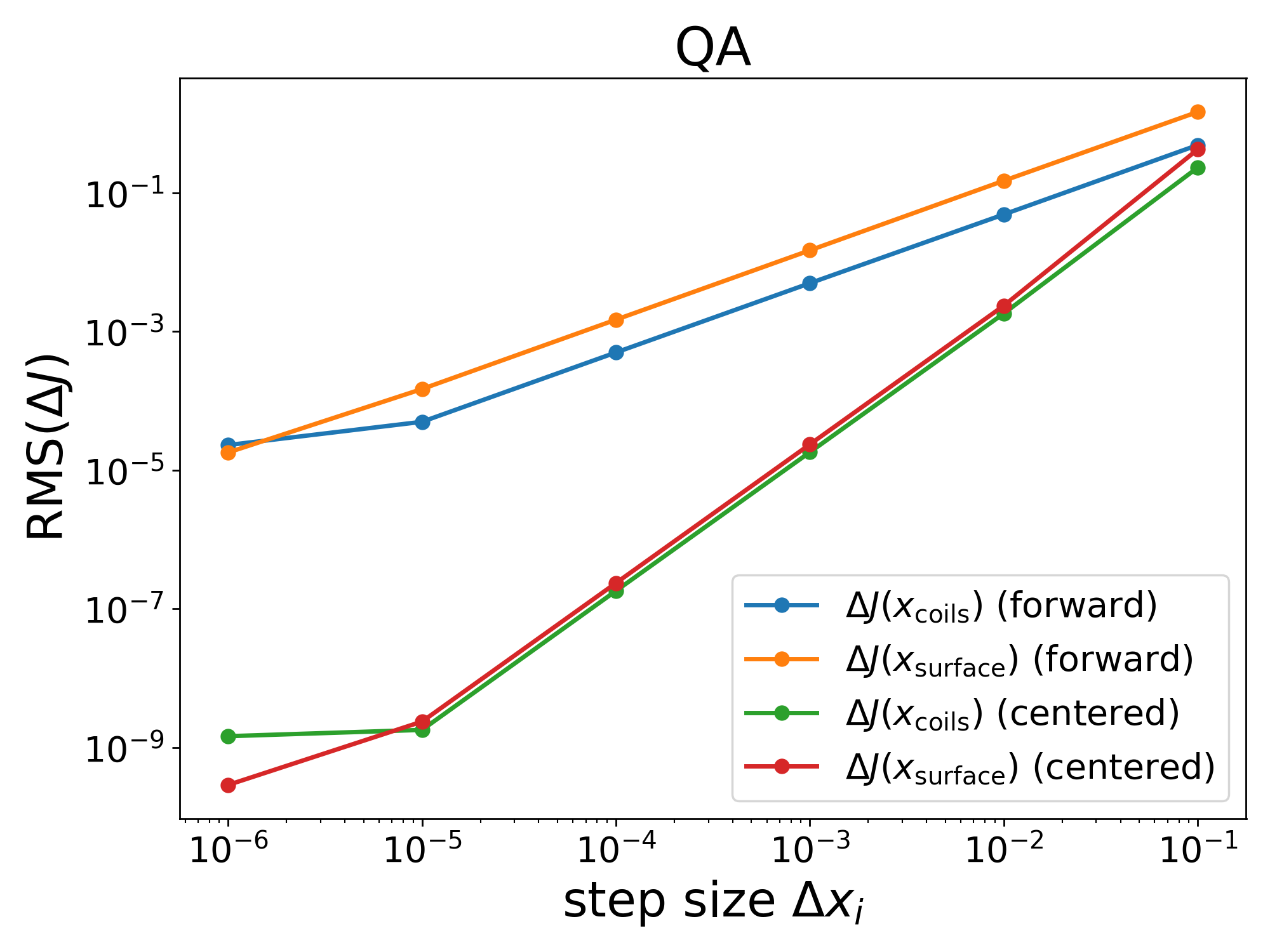}
    \includegraphics[width=.49\textwidth]{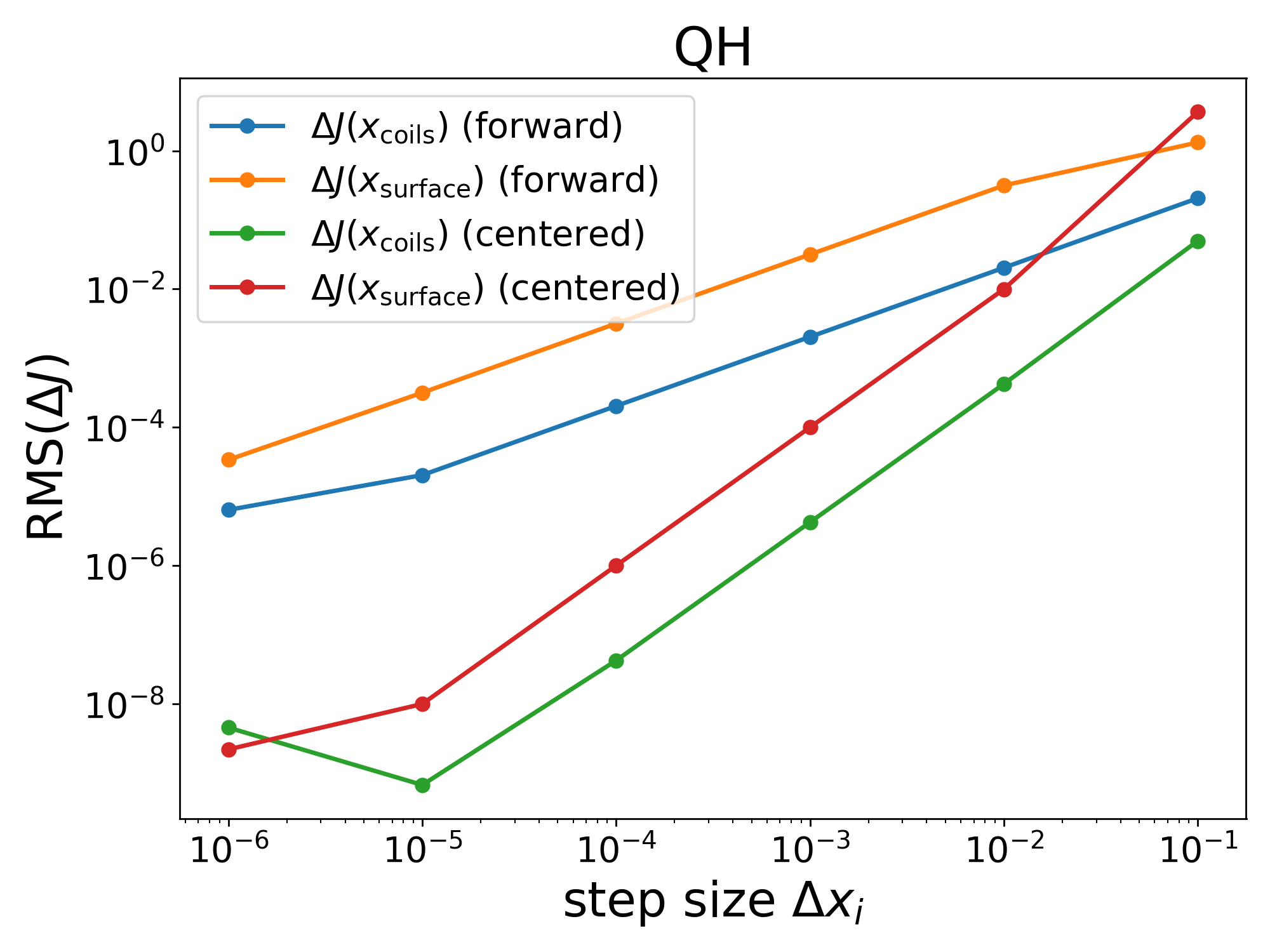}
    \caption{First-order (forward) and second-order (centered) convergence shown using the root mean square (RMS) of the difference between the gradient vector computed using one of \cref{eq:estimate_dJ2,eq:forward} and the gradient vector returned by SIMSOPT, both for the precise QA and QH cases.}
    \label{fig:estimate_dJ}
\end{figure}

We then perform a convergence test on the combined objective function $J$.
This is done to ensure that the implementation of finite differences in the derivatives of $J_1$ and the addition operation performed in \cref{eq:J} still allows the gradient of $J$ to be estimated accurately.
For this purpose, we apply the centered finite difference formula in \cref{eq:estimate_dJ2} to $J$ by replacing $J_2$ with $J$ and $\mathbf x$ with $\mathbf x_{\text{coils}}$ or $\mathbf x_{\text{surface}}$.
For comparison, we also estimate the gradients using a forward finite difference formula
\begin{equation}
    \frac{\Delta J}{\Delta x} = \frac{J(\mathbf x + \mathbf h \Delta x)-J(\mathbf x)}{\Delta x},
\label{eq:forward}
\end{equation}
and verify both first and second-order convergence using forward and centered finite differences, respectively.
We show in \cref{fig:estimate_dJ} the root mean square (RMS) of the difference between the gradient vector computed using one of \cref{eq:estimate_dJ2,eq:forward} and the gradient vector returned by SIMSOPT, both for the precise QA and QH cases.
The expected first and second-order convergence is found both for the gradient with respect to $\mathbf x_{\text{coils}}$ and $\mathbf x_{\text{surface}}$.

\section{Numerical Results}
\label{sec:results}

We now show the results found using the single-stage approach proposed here by adding it as a third step in the stellarator optimization process.
Namely, we first run stage 1 and stage 2 optimizations where $J_1$ and $J_2$ are minimized sequentially.
Then, we run a single-stage optimization to obtain a fixed boundary equilibrium that faithfully reproduces the magnetic field stemming from the external coils and minimizes $J$ in \cref{eq:J}.
{In this process, it is worth mentioning that, for each function evaluation, the stage 1 quantities (related to the equilibrium part) take significantly longer to evaluate than the stage 2 quantities.}

We note that all numerical results presented in this work pertain to vacuum fields.
However, we emphasize that our approach is readily extendable to finite beta scenarios.
By considering vacuum fields first, we can employ Poincare plots as a sensitive diagnostic tool to verify the consistency of our results.
This not only enables a more rigorous evaluation of our method but also provides a solid foundation for future applications to more complex scenarios involving finite plasma pressure.

In order to assess if the fixed boundary equilibrium faithfully reproduces the magnetic field stemming from the external coils, we create a quadratic flux minimizing surface (QFM) from the resulting coils \cite{Dewar1994} with the same volume as the fixed boundary surface $S$ and compare it to the original equilibrium.
These are surfaces that minimize
\begin{equation}
    f(S)=\frac{\int_0^{2\pi} \int_0^{2\pi/n_{\text{fp}}} (\mathbf B \cdot \mathbf n)^2 d \theta d \varphi}{\int_0^{2\pi} \int_0^{2\pi/n_{\text{fp}}} B^2 d \theta d \varphi} + \frac{1}{2} [\text{Vol}(\text{QFM})-\text{Vol}(S)]^2,
\end{equation}
without constraints on the angles that parametrize the surface, $\text{Vol}$(S) is the total volume within the fixed-boundary surface $S$ and $\text{Vol}(\text{QFM})$ the volume within the QFM surface.
Finally, we also verify our results by comparing the QFM and fixed-boundary surfaces with Poincaré plots, which are cross-sections of the magnetic field lines traced using the Biot-Savart magnetic field from the resulting coils.
It is worth noting that for vacuum fields, we have observed that running fixed-boundary VMEC inside a QFM surface results in better accuracy than running free-boundary VMEC, based on comparisons of flux surface shapes to Poincare plots.
Consequently, our analysis of the final configurations is based on QFM surfaces instead of free-boundary VMEC.

\subsection{Four-Coil Stellarator}
\label{sec:simple}

We first apply this method to find a set of four simplified coils, two circular coils at the top and bottom of the device, and two interlinking coils.
Similar configurations include Columbia Non-neutral Torus (CNT) device \cite{Pedersen2002}, a stellarator experiment at Columbia University with four circular coils which are used here as the starting point for the optimization, and the Compact Stellarator with Simple Coils (CSSC) \cite{Yu2022}.
First, we perform an optimization with the top and bottom coils fixed, while the two symmetric interlinking coils are allowed to vary.
Then, we perform a second optimization adding the degrees of freedom of the top coil so that it is non-circular, with the bottom coil symmetric to the top one.

As an optimization goal, we aim at a quasi-axisymmetric device with an aspect ratio $A=3.5$ and mean rotational transform of $\iota=0.23$.
As coil parameters, we let each of the independent coils have a total of $N_F=7$ Fourier modes, the interlinking coils to have $L_{\text{max}}=3.8$, $\kappa_{\text{max}}=12$, $\kappa_{\text{msc}}=12$ and the top and bottom coils to have $L_{\text{max}}=7.0$, $\kappa_{\text{max}}=3.5$, and $\kappa_{\text{msc}}=4.5$ with a minimum distance between coils of 0.15.
As surface parameters, we use a total of $M_\text{pol}=N_\text{tor}=3$ surface Fourier modes.

\begin{figure}
    \centering
    \begin{minipage}{0.99\linewidth}
            \begin{subfigure}{}
                \includegraphics[width=0.99\textwidth,
                ]{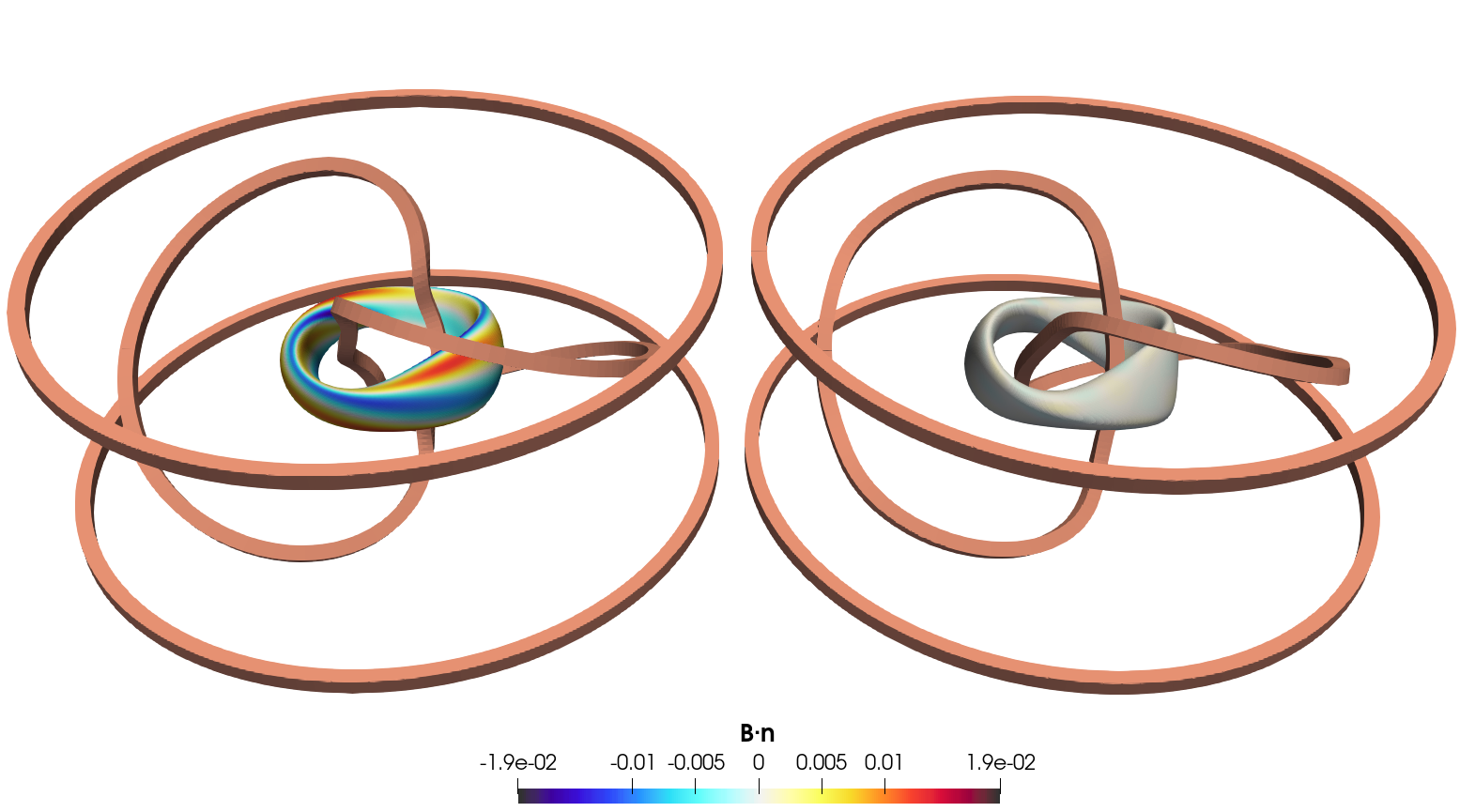}
            \end{subfigure}
        \end{minipage}
    \begin{minipage}{.56\linewidth}
            \begin{subfigure}{}
                \includegraphics[width=\textwidth,trim={0.0cm 0.0cm 0.0cm 0.0cm},clip]{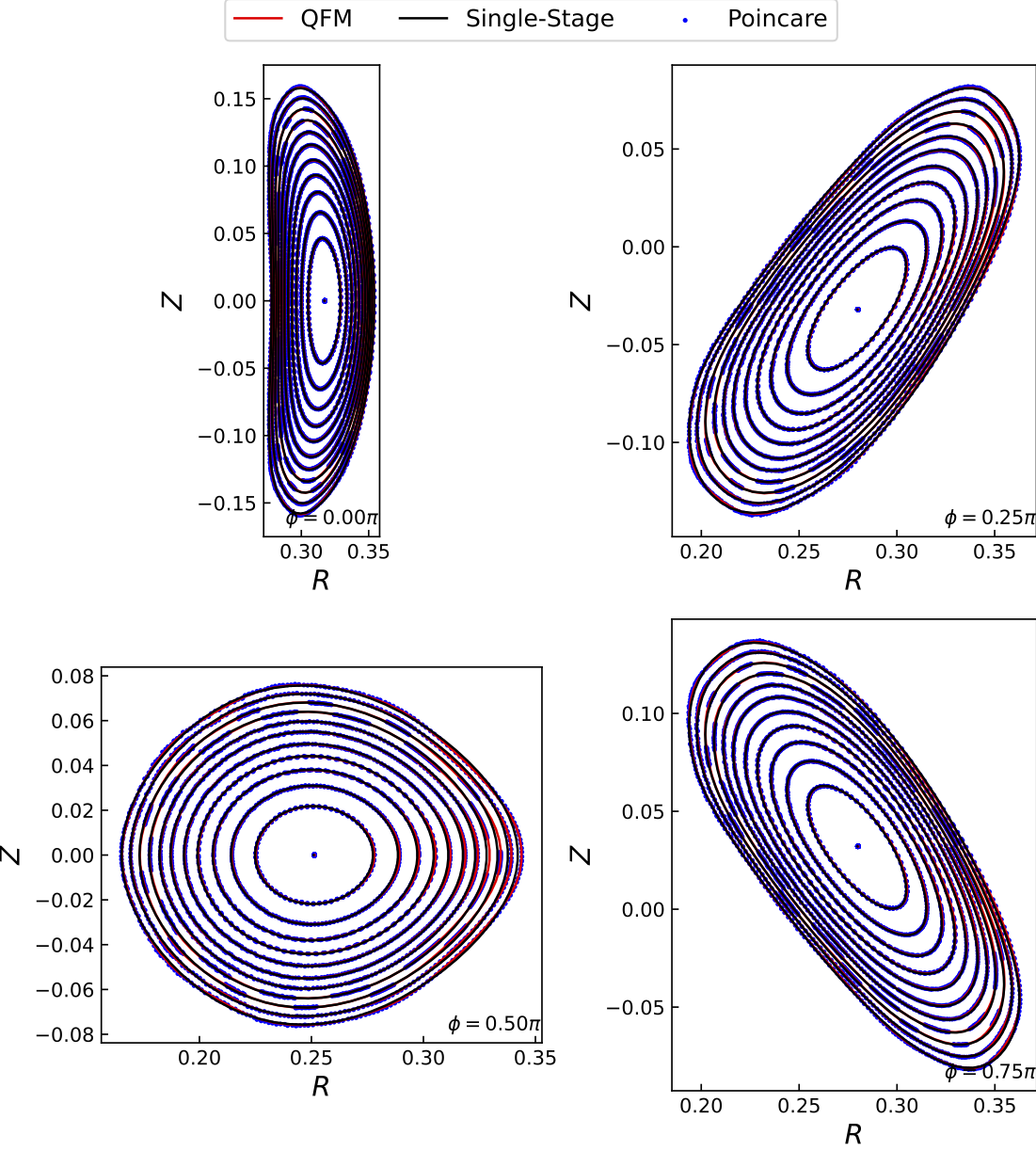}
            \end{subfigure}
        \end{minipage}
    \begin{minipage}{.43\linewidth}
        \begin{subfigure}{}
            \includegraphics[width=\textwidth]{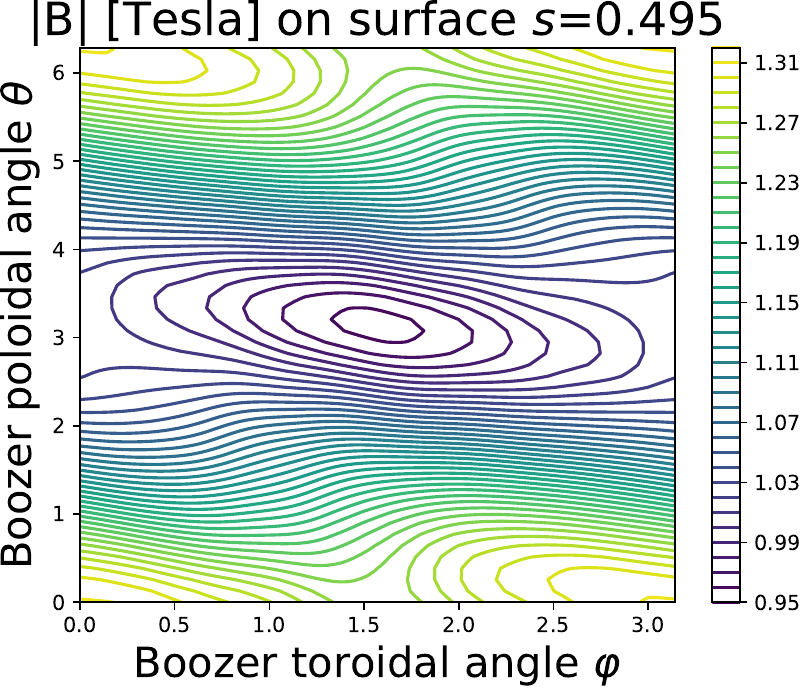}
        \end{subfigure} \\
        \begin{subfigure}{}
            \includegraphics[width=\textwidth,trim={0.0cm 12.0cm 22.5cm 0.5cm},clip]{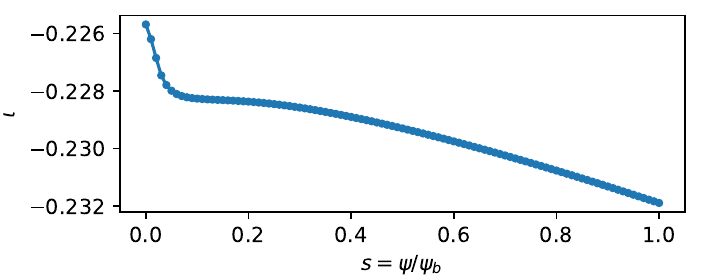}
        \end{subfigure} 
    \end{minipage}
    \caption{Optimization with circular top and bottom coils.
    Top: standard stellarator optimization approach where stage 1 and stage 2 optimizations were performed sequentially (left) and the single-stage optimization result (right).
    Lower Left: Superposition of magnetic surfaces at constant cylindrical toroidal angle $\phi$ of the QFM and the final single-stage equilibrium, as well as the Poincaré plot resulting from tracing magnetic field lines in the obtained coils. Middle Right: Contours of constant magnetic field strength on a surface at $s=0.495$ in Boozer coordinates $(\theta, \varphi)$. Bottom Right: profile of rotational transform $\iota$.
    \label{fig:properties_CNT_circular}
    }
\end{figure}

We show in \cref{fig:properties_CNT_circular} the result of the first optimization with fixed circular top and bottom coils.
The configuration associated with the stage 1 and stage 2 independent optimizations is shown in \cref{fig:properties_CNT_circular} (top left) where the residuals of the quasisymmetry objective function $f_{\text{QS}}$ of stage 1 is $5.3 \times 10^{-5}$, and the squared flux $f_{\text{QF}}$ of stage 2 is $6.3 \times 10^{-4}$.
The resulting single-stage optimization is shown in \cref{fig:properties_CNT_circular} (top right) with corresponding residuals of $f_{\text{QS}}=1.7 \times 10^{-2}$ and a squared flux of $f_{\text{QF}}=1.6 \times 10^{-6}$.
We then run VMEC in fixed-boundary mode using the QFM surfaces obtained from the stage 1 and stage 2 independent optimizations and the single-stage approach.
In this case, the first yields a quasisymmetry objective function of $f_{\text{QS}}=3.0 \times 10^{-2}$ while the second yields $f_{\text{QS}}=1.7 \times 10^{-2}$, the same value as the optimization result.
The second $f_{\text{QS}}$ number is smaller than the first, indicating the effectiveness of the stage-3 optimization.
This shows that while the stage 1 optimization can result in very precise quasisymmetric (or quasi-isodynamic) configurations, the resulting configuration from a stage 2 optimization may not retain the same expected properties.
The figure in \cref{fig:properties_CNT_circular} (lower left) shows that the minimization of the squared flux leads to an agreement between the single-stage fixed boundary equilibrium, a fixed boundary equilibrium based on the QFM surface, and the Poincaré plots.
The contours of constant magnetic field at the $s=0.495$ surface resulting from a VMEC run based on the QFM surface obtained using the final coil configuration in Boozer coordinates, which assess the degree of quasisymmetry associated with this configuration, and its rotational transform profile, are shown in \cref{fig:properties_CNT_circular} (middle and bottom right).

\begin{figure}
    \centering
    \begin{minipage}{1.0\linewidth}
            \begin{subfigure}{}
                \includegraphics[width=1.0\textwidth,
                ]{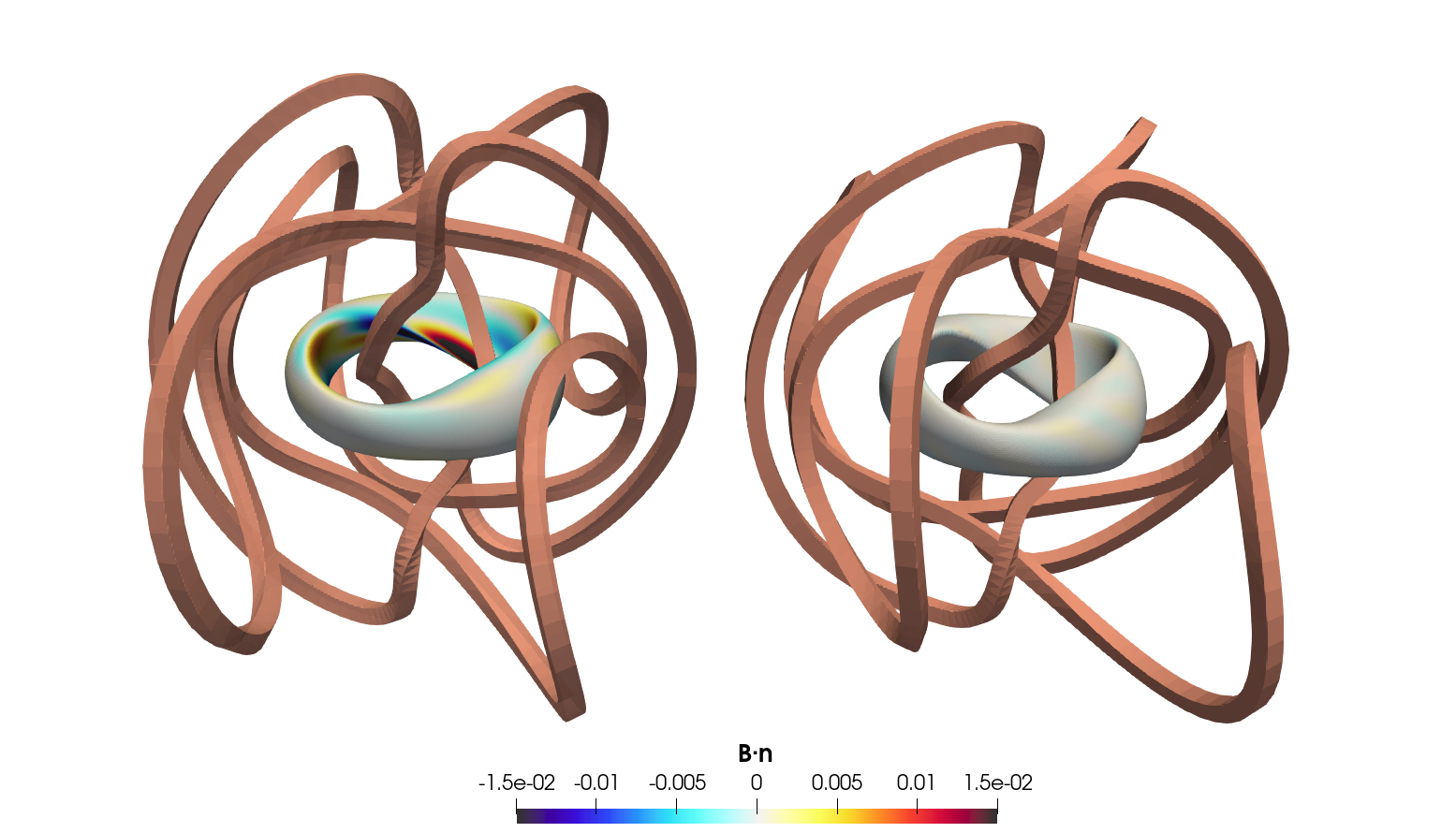}
            \end{subfigure}
        \end{minipage}
    \begin{minipage}{.56\linewidth}
            \begin{subfigure}{}
                \includegraphics[width=\textwidth]{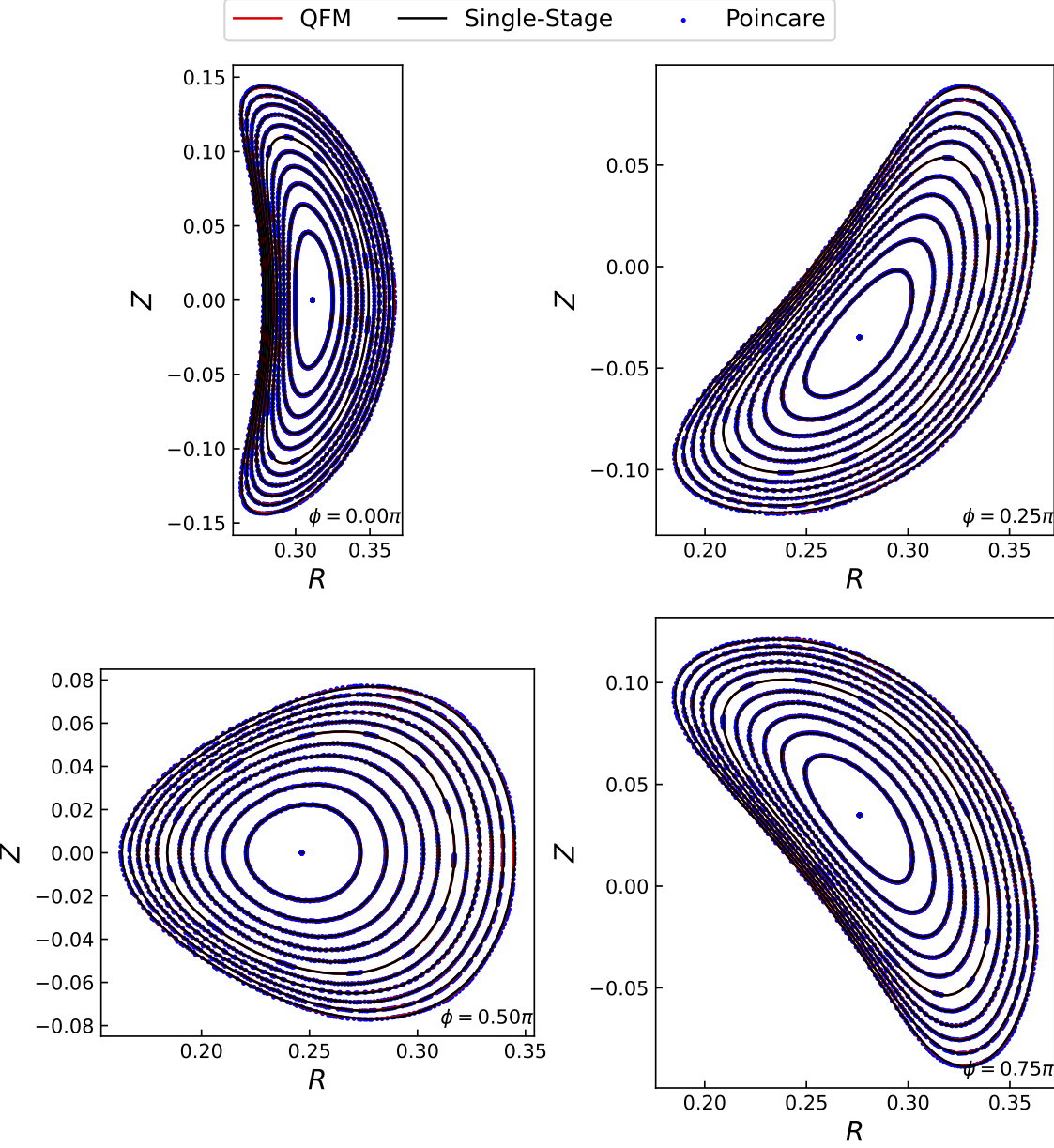}
            \end{subfigure}
        \end{minipage}
    \begin{minipage}{.43\linewidth}
        \begin{subfigure}{}
            \includegraphics[width=\textwidth]{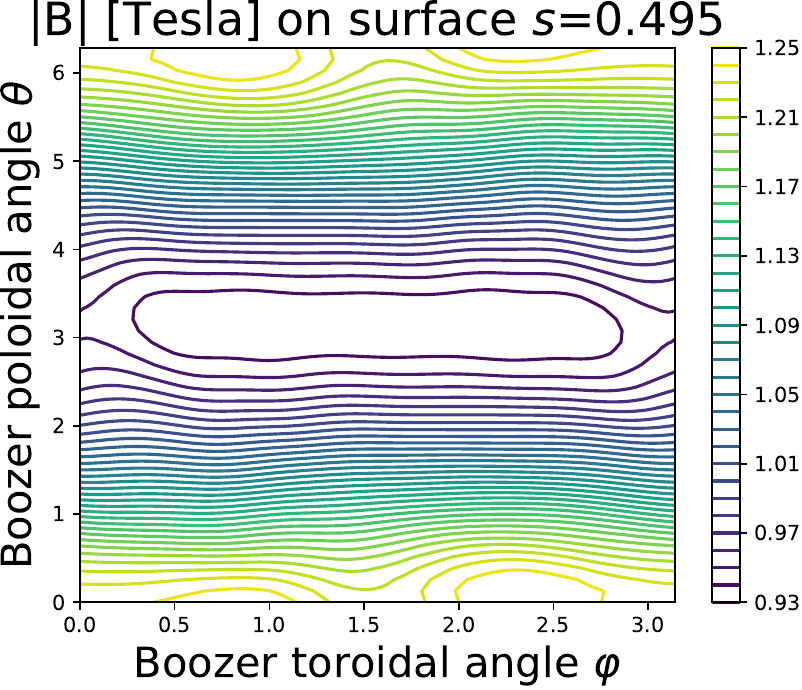}
        \end{subfigure} \\
        \begin{subfigure}{}
            \includegraphics[width=\textwidth,trim={0.0cm 12.0cm 23.0cm 0.5cm},clip]{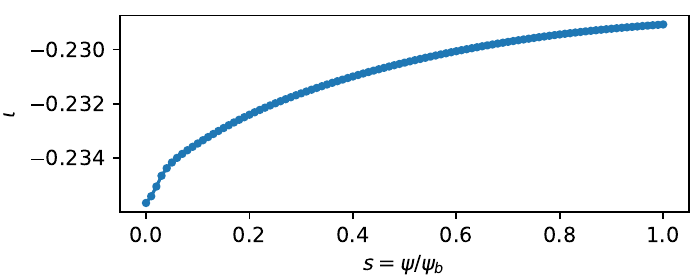}
        \end{subfigure} 
    \end{minipage}
    \caption{Optimization with free top and bottom coils.
    Top: standard stellarator optimization approach where stage 1 and stage 2 optimizations were performed sequentially (left) and the single-stage optimization result (right).
    Lower Left: Superposition of magnetic surfaces at constant cylindrical toroidal angle $\phi$ of the QFM and the final single-stage equilibrium, as well as the Poincaré plot resulting from tracing magnetic field lines in the obtained coils. Middle Right: Contours of constant magnetic field strength on a surface at $s=0.495$ in Boozer coordinates $(\theta, \varphi)$. Bottom Right: profile of rotational transform $\iota$.
    \label{fig:properties_CNT}
    }
\end{figure}

We show in \cref{fig:properties_CNT} the result of the second optimization with free top and bottom coils.
The configuration associated with the stage 1 and stage 2 independent optimizations is shown in \cref{fig:properties_CNT} (top left) where the residuals of the quasisymmetry objective function $f_{\text{QS}}$ of stage 1 is $5.3 \times 10^{-5}$, and the squared flux $f_{\text{QF}}$ of stage 2 is $8.7 \times 10^{-5}$.
The resulting single-stage optimization is shown in \cref{fig:properties_CNT} (top right) with corresponding residuals of $f_{\text{QS}}=4.9 \times 10^{-3}$ and a squared flux of $f_{\text{QF}}=9.3 \times 10^{-6}$.
We then run VMEC in fixed-boundary mode using the QFM surfaces obtained from the stage 1 and stage 2 independent optimizations and the single-stage approach.
In this case, the first yields a quasisymmetry objective function of $f_{\text{QS}}=1.8 \times 10^{-2}$ while the second yields $f_{\text{QS}}=5.0 \times 10^{-3}$.
The second value is smaller than the former, again showing that the combined plasma-and-coils optimization provides a better result than the traditional two-stage method.
The figure in \cref{fig:properties_CNT} (lower left) shows that the minimization of the squared flux leads to an agreement between the single-stage fixed boundary equilibrium, a fixed boundary equilibrium based on the QFM surface, and the Poincaré plots.
The contours of constant magnetic field at the $s=0.495$ surface in Boozer coordinates, which assess the degree of quasisymmetry associated with this configuration, and its rotational transform profile are shown in \cref{fig:properties_CNT} (middle and bottom right).

Comparison of \cref{fig:properties_CNT_circular,fig:properties_CNT} reveals that the degree of quasisymmetry can be significantly improved by allowing for variation in the Helmholtz coils.
However, this comes at the cost of increased complexity in the coil shapes, as demonstrated by the top row of \cref{fig:properties_CNT}.

\subsection{Quasi-axisymmetry}
\label{sec:QA}

\begin{figure}
    \centering
    \begin{minipage}{1.\linewidth}
        \begin{subfigure}{}
            \includegraphics[width=1.\textwidth,
            ]{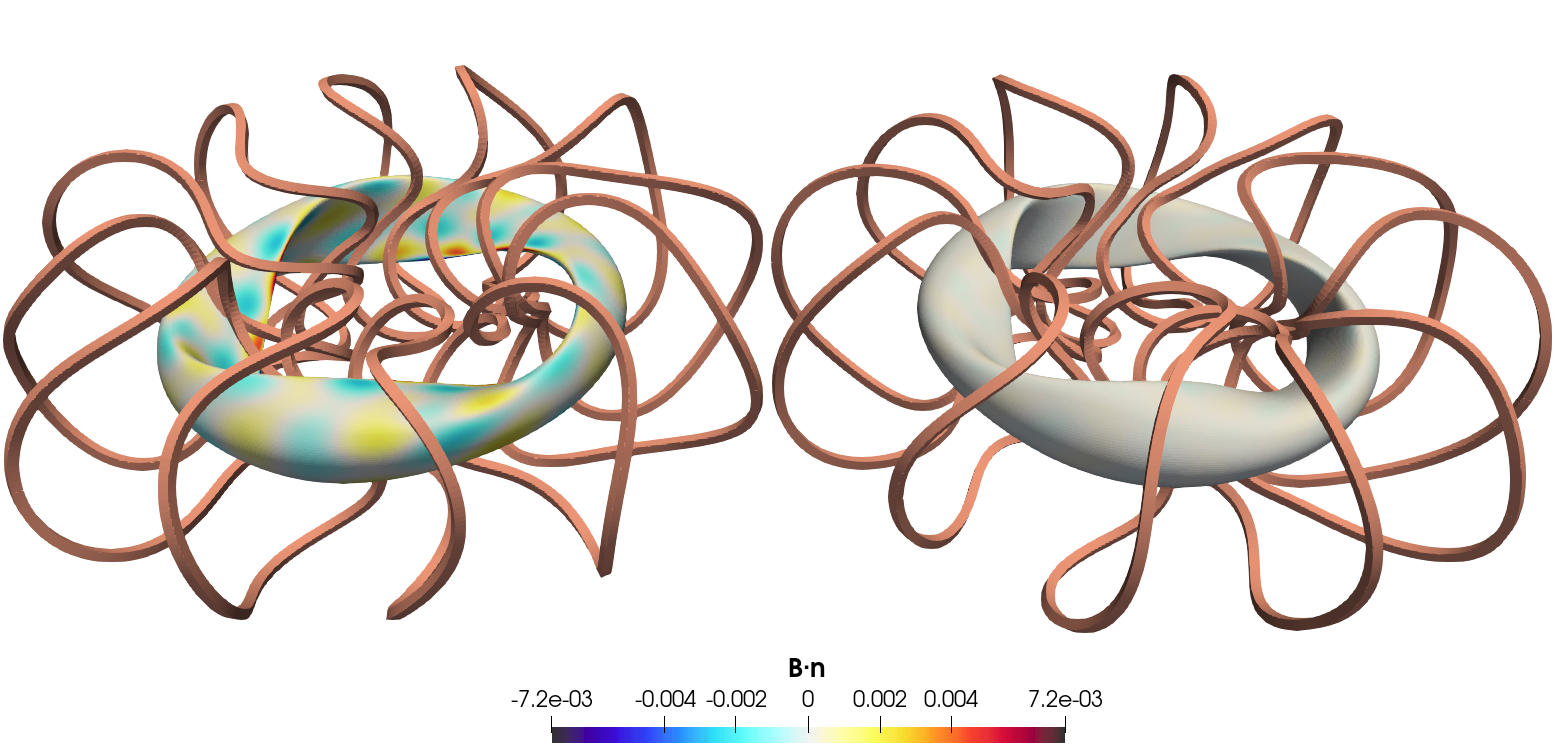}
        \end{subfigure}
    \end{minipage}
    \begin{minipage}{.56\linewidth}
            \begin{subfigure}{}
                \includegraphics[width=\textwidth]{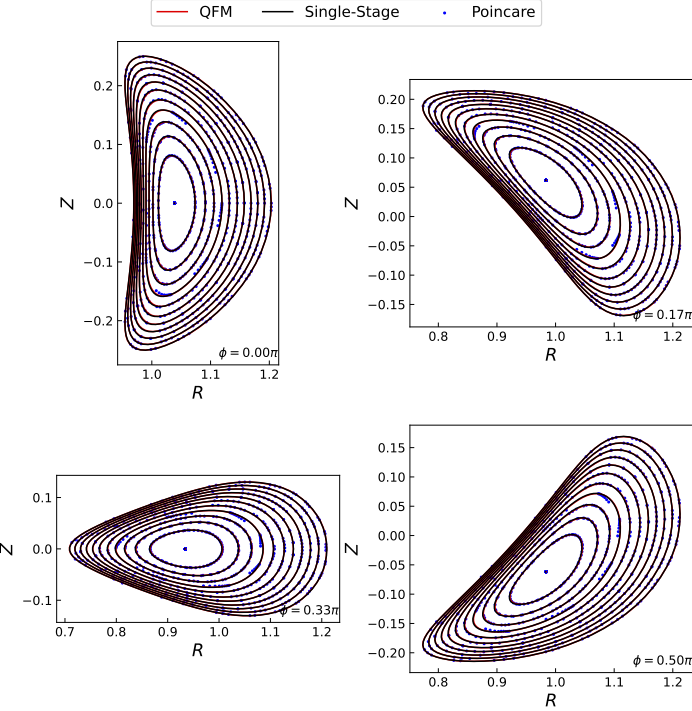}
            \end{subfigure}
        \end{minipage}
    \begin{minipage}{.43\linewidth}
        \begin{subfigure}{}
            \includegraphics[width=\textwidth]{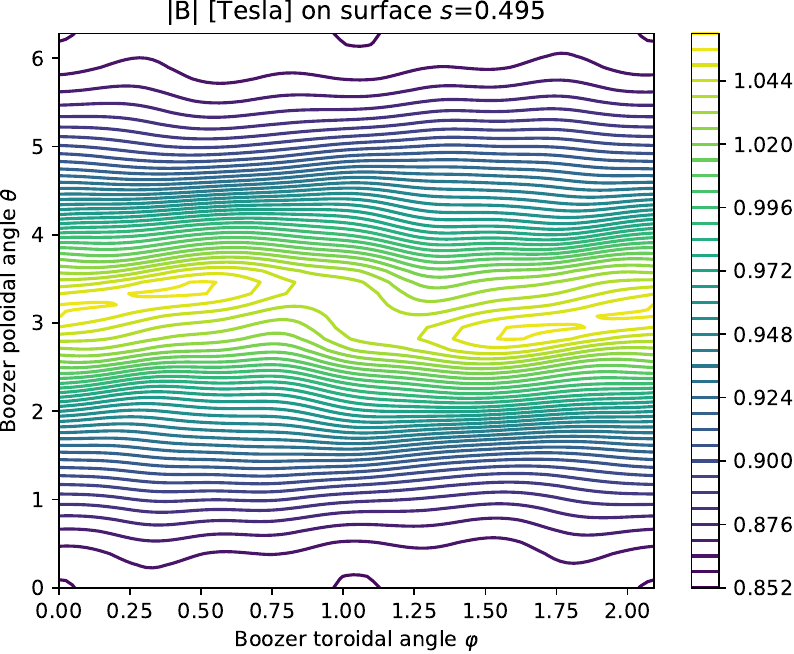}
        \end{subfigure} \\
        \begin{subfigure}{}
            \includegraphics[width=\textwidth,trim={0.0cm 12.0cm 23.0cm 0.5cm},clip]{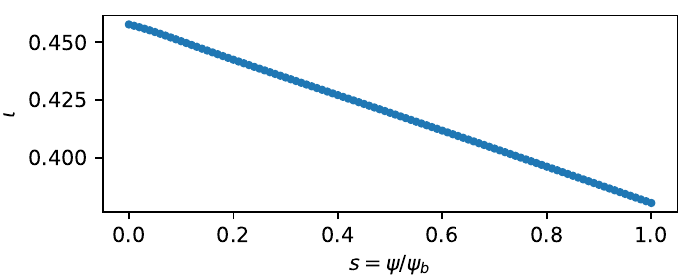}
        \end{subfigure} 
    \end{minipage}
    \caption{Quasi-axisymmetric stellarator with 3 field periods and 2 coils per half-field period.
    Top: standard stellarator optimization approach where stage 1 and stage 2 optimizations were performed sequentially (left) and the single-stage optimization result (right). 
    Lower Left: Superposition of magnetic surfaces at constant cylindrical toroidal angle $\phi$ of the QFM and the final single-stage equilibrium, as well as the Poincaré plot resulting from tracing magnetic field lines in the obtained coils.
    Middle Right: Contours of constant magnetic field strength on a surface at $s=0.495$ in Boozer coordinates $(\theta, \varphi)$. Bottom Right: profile of rotational transform $\iota$.
    \label{fig:properties_QAncoils2nfp3}
    }
\end{figure}

We now optimize a three-field period quasi-axisymmetric stellarator with 2 coils per half-field period.
As an optimization goal, we aim at a quasi-axisymmetric device with an aspect ratio $A=6.0$ and mean rotational transform of $\iota=0.42$.
As coil parameters, we let each of the independent coils have a total of $N_F=16$ Fourier modes, each coil to have $L_{\text{max}}=5.5$, $\kappa_{\text{max}}=5.0$, and $\kappa_{\text{msc}}=5.0$ with a minimum distance between coils of 0.1.
The configuration associated with the stage 1 and stage 2 independent optimizations is shown in \cref{fig:properties_QAncoils2nfp3} (top left) where the residuals of the quasisymmetry objective function $f_{\text{QS}}$ of stage 1 is $2.3 \times 10^{-4}$, and the squared flux $f_{\text{QF}}$ of stage 2 is $9.8 \times 10^{-5}$.
The resulting single-stage optimization is shown in \cref{fig:properties_QAncoils2nfp3} (top right) with corresponding residuals of $f_{\text{QS}}=9.3 \times 10^{-3}$ and a squared flux of $f_{\text{QF}}=7.9 \times 10^{-6}$.
We then run VMEC in fixed-boundary mode using the QFM surfaces obtained from the stage 1 and stage 2 independent optimizations and the single-stage approach.
In this case, the first yields a quasisymmetry objective function of $f_{\text{QS}}=1.7 \times 10^{-2}$ while the second yields $f_{\text{QS}}=9.1 \times 10^{-3}$.
These values show that once again, better quasisymmetry is obtained using the combined plasma-and-coils optimization compared to consecutive stage 1 and stage 2 optimization.
The figure in \cref{fig:properties_QAncoils2nfp3} (lower left) shows that the minimization of the squared flux leads to an agreement between the single-stage fixed boundary equilibrium, a fixed boundary equilibrium based on the QFM surface, and the Poincaré plots.
The contours of constant magnetic field at the $s=0.495$ surface in Boozer coordinates, which assess the degree of quasisymmetry associated with this configuration, and its rotational transform profile are shown in \cref{fig:properties_QAncoils2nfp3} (middle and bottom right).

\subsection{Quasi-Helical Symmetry}
\label{sec:QH}

\begin{figure}
    \centering
    \begin{minipage}{1.\linewidth}
        \begin{subfigure}{}
            \includegraphics[width=1.0\textwidth,
            ]{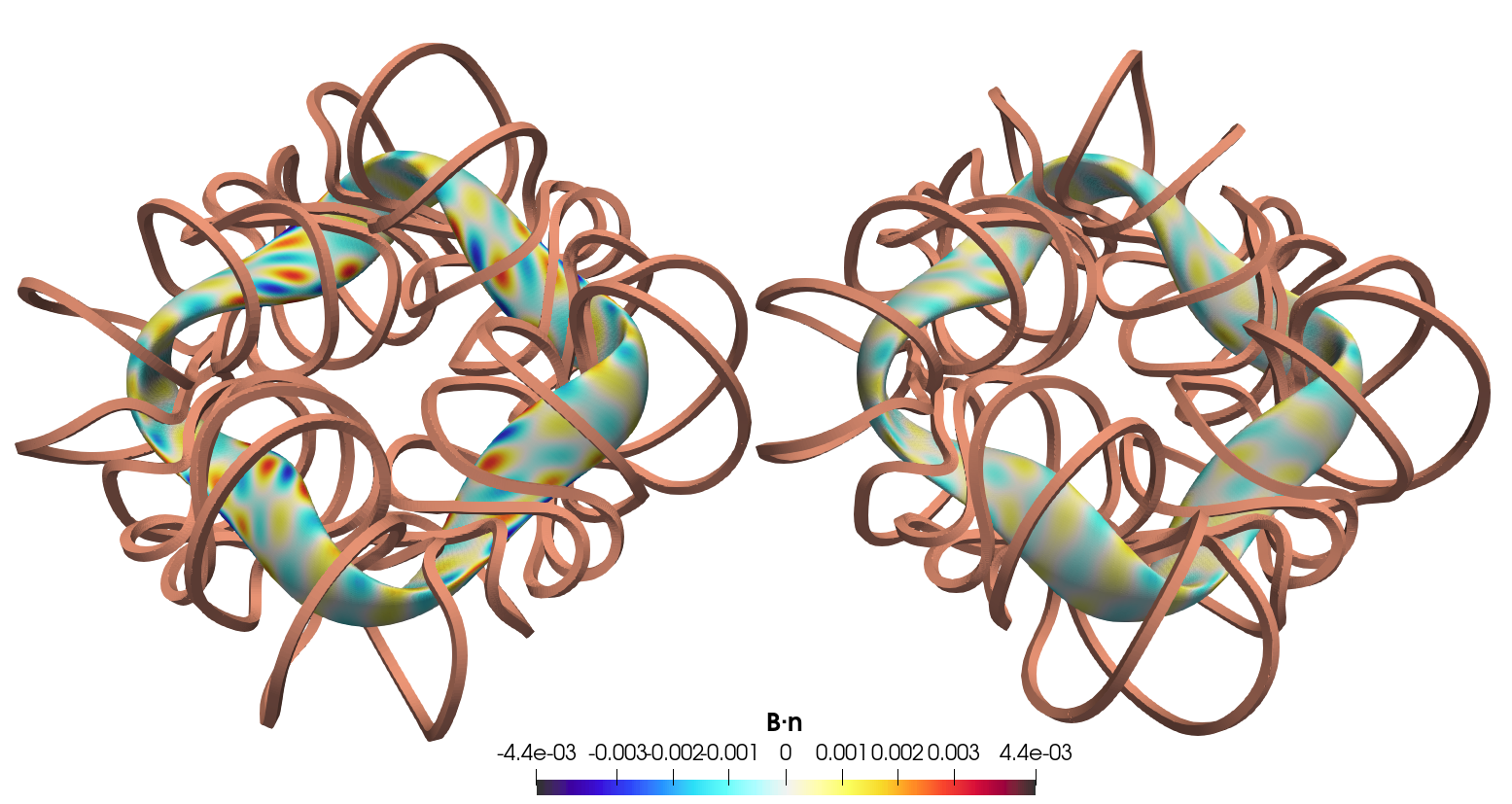}
        \end{subfigure}
    \end{minipage}
    \begin{minipage}{.55\linewidth}
            \begin{subfigure}{}
                \includegraphics[width=\textwidth]{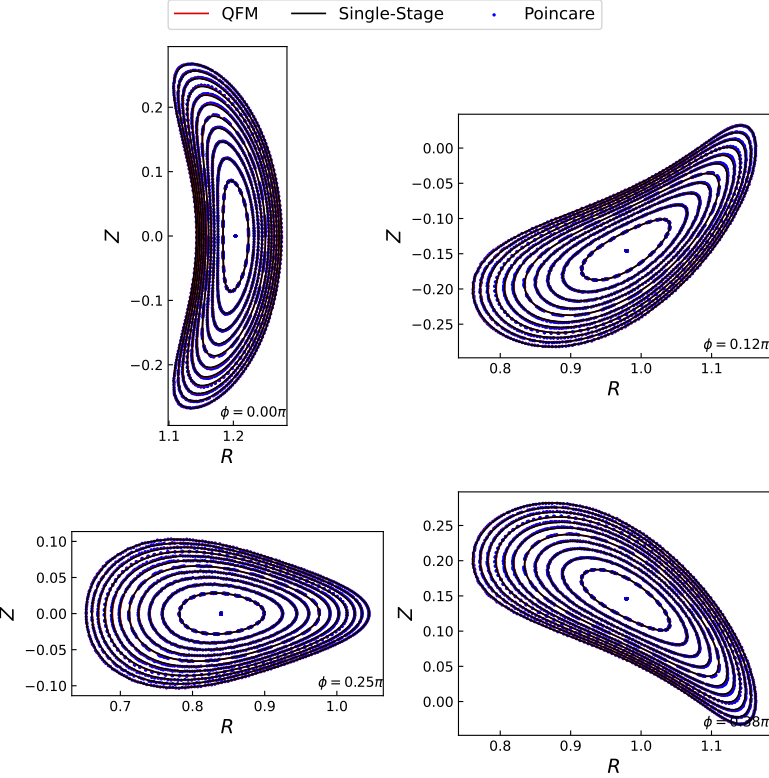}
            \end{subfigure}
        \end{minipage}
    \begin{minipage}{.43\linewidth}
        \begin{subfigure}{}
            \includegraphics[width=\textwidth]{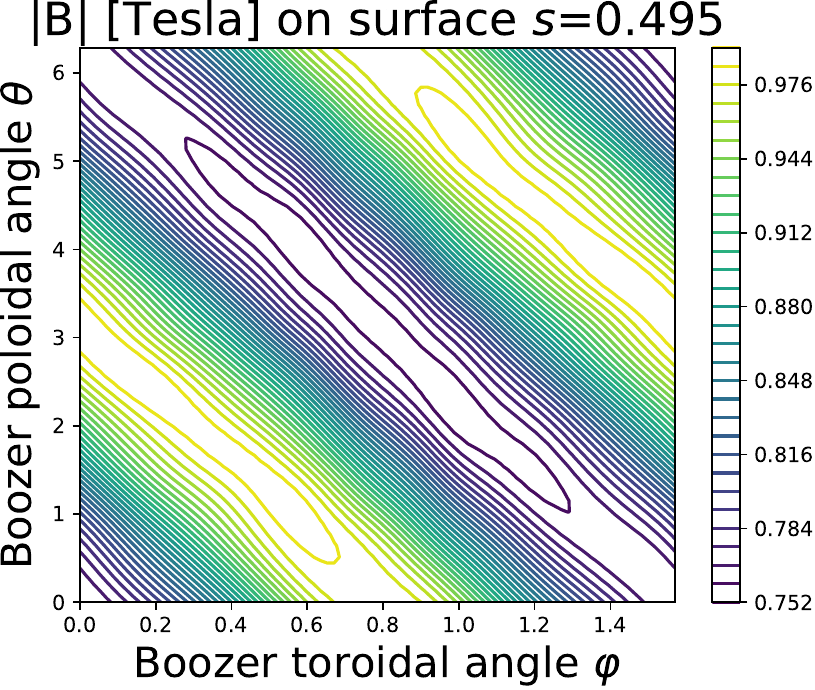}
        \end{subfigure} \\
        \begin{subfigure}{}
            \includegraphics[width=\textwidth,trim={0.0cm 12.0cm 23.0cm 0.5cm},clip]{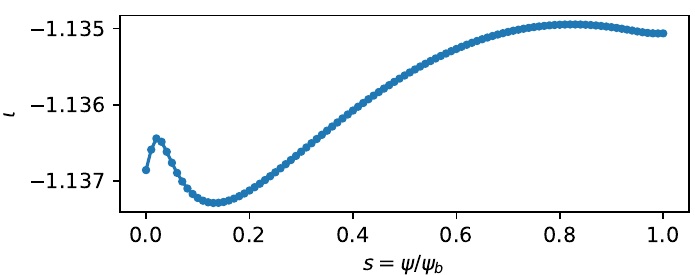}
        \end{subfigure} 
    \end{minipage}
    \caption{Quasi-helical symmetric stellarator with 4 field periods and 3 coils per half-field period.
    Top: standard stellarator optimization approach where stage 1 and stage 2 optimizations were performed sequentially (left) and the single-stage optimization result (right).
    Lower Left: Superposition of magnetic surfaces at constant cylindrical toroidal angle $\phi$ of the QFM and the final single-stage equilibrium, as well as the Poincaré plot resulting from tracing magnetic field lines in the obtained coils.
    Middle Right: Contours of constant magnetic field strength on a surface at $s=0.495$ in Boozer coordinates $(\theta, \varphi)$. Bottom Right: profile of rotational transform $\iota$.
    \label{fig:properties_QHncoils3nfp4}
    }
\end{figure}

We present a systematic optimization of a four-field period quasi-helical symmetric stellarator. The goal of the optimization was to achieve a device with an aspect ratio of $A=7.0$ that displays quasi-helical symmetry.
To achieve this, we utilized 3 independent coils, each of which contained $N_F=16$ Fourier modes. The maximum length of each coil was set at $L_{\text{max}}=3.5$ while the maximum value of the shaping parameter $\kappa$ was set at $\kappa_{\text{max}}=10.0$, and $\kappa_{\text{msc}}=10.0$ with a minimum distance between coils of 0.08.
The results of stage 1 and stage 2 consecutive optimizations are displayed in \cref{fig:properties_QHncoils3nfp4} (top left), where the residuals of the quasisymmetry objective function $f_{\text{QS}}$ is shown to be $2.0 \times 10^{-3}$ and the squared flux $f_{\text{QF}}$ is $3.1 \times 10^{-5}$. The final single-stage optimization is shown in \cref{fig:properties_QHncoils3nfp4} (top right) with corresponding residuals of $f_{\text{QS}}=1.3 \times 10^{-2}$ and a squared flux of $f_{\text{QF}}=8.3 \times 10^{-6}$.
We then run VMEC in fixed-boundary mode using the QFM surfaces obtained from the stage 1 and stage 2 consecutive optimizations and the single-stage approach.
In this case, the first yields a quasisymmetry objective function of $f_{\text{QS}}=5.3 \times 10^{-2}$ while the second yields $f_{\text{QS}}=2.0 \times 10^{-2}$.
As demonstrated in \cref{fig:properties_QHncoils3nfp4} (lower left), the minimization of the squared flux results in consistency between the single-stage fixed boundary equilibrium, a fixed boundary equilibrium based on the QFM surface, and the Poincaré plots. Additionally, the contours of constant magnetic field at the $s=0.495$ surface in Boozer coordinates, which assess the degree of quasisymmetry in the configuration, and its rotational transform profile are displayed in \cref{fig:properties_QHncoils3nfp4} (middle and bottom right).
This optimization provides an important contribution to the field of stellarator optimization. By systematically obtaining coils for a quasi-helical symmetric stellarator, we have taken a step towards improving the performance and stability of these devices in fusion energy applications.

\subsection{Quasi-isodynamic}
\label{sec:QI}

\begin{figure}
    \centering
    \begin{minipage}{1.\linewidth}
        \begin{subfigure}{}
            \includegraphics[width=.95\textwidth,
            ]{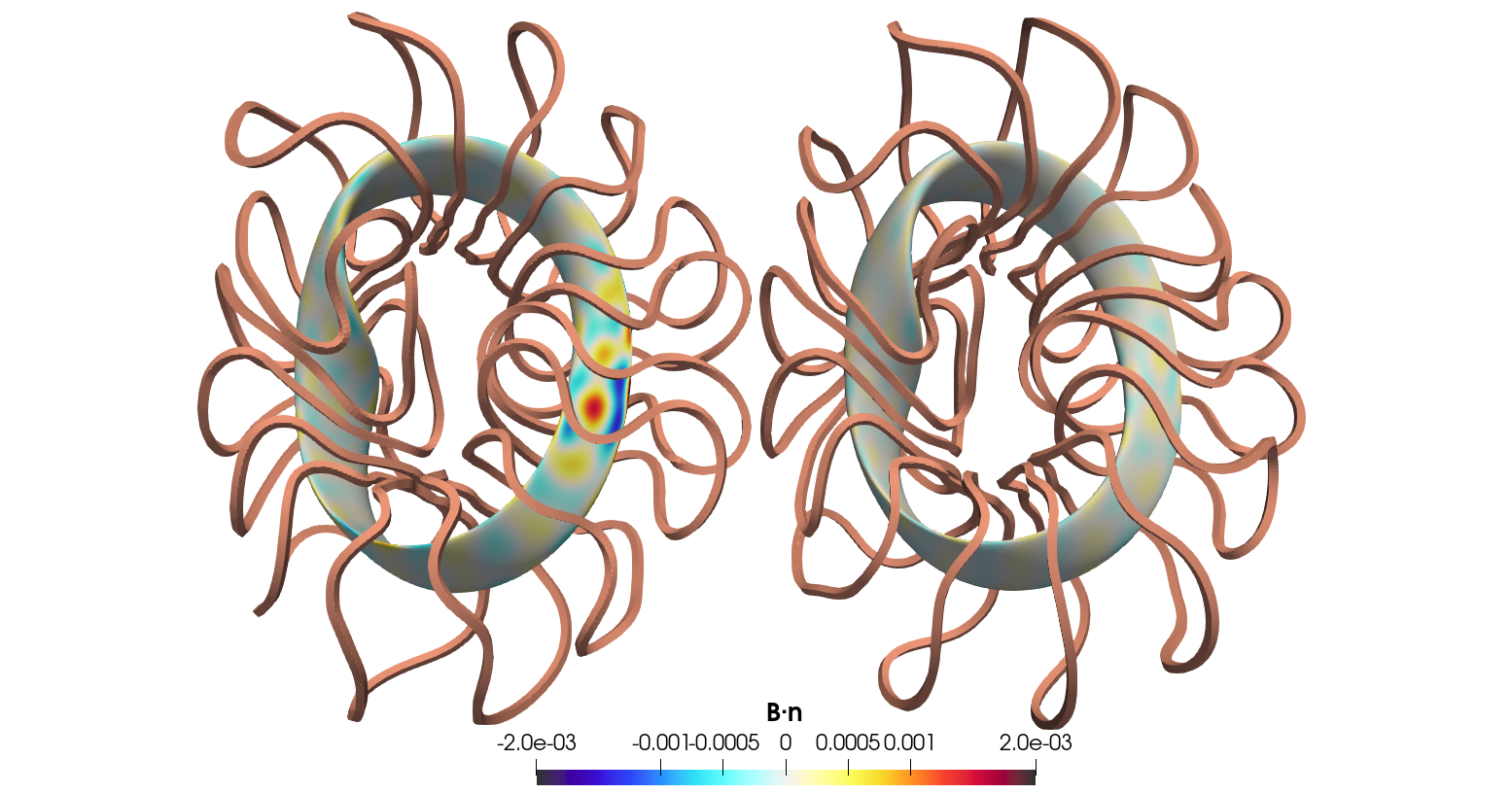}
        \end{subfigure}
    \end{minipage}
    \begin{minipage}{.56\linewidth}
            \begin{subfigure}{}
                \includegraphics[width=\textwidth]{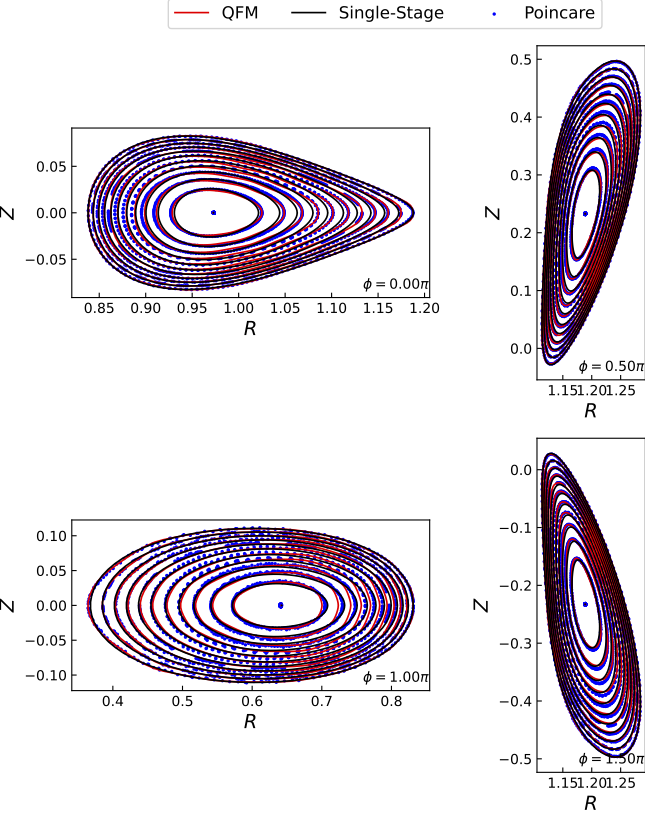}
            \end{subfigure}
        \end{minipage}
    \begin{minipage}{.43\linewidth}
        \begin{subfigure}{}
            \includegraphics[width=\textwidth]{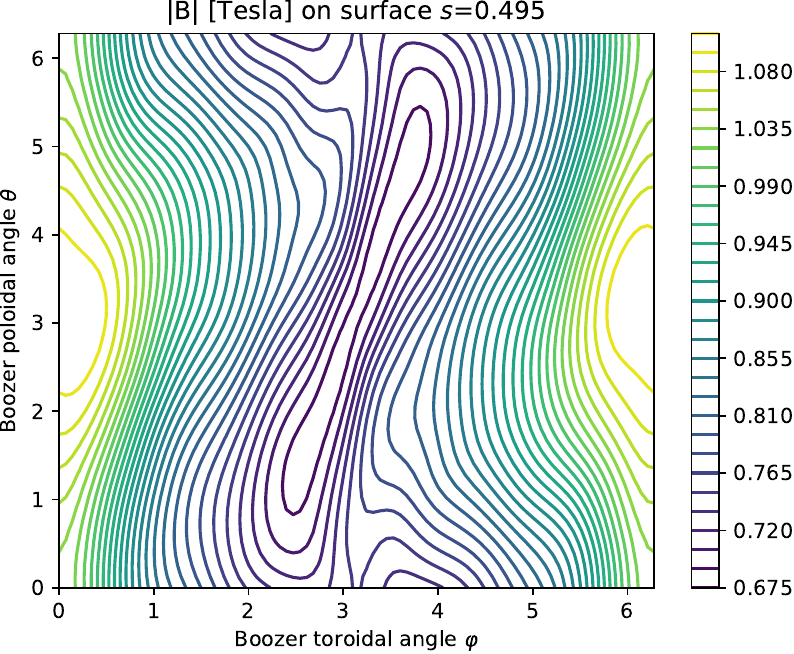}
        \end{subfigure} \\
        \begin{subfigure}{}
            \includegraphics[width=\textwidth,trim={0.0cm 12.0cm 23.0cm 0.5cm},clip]{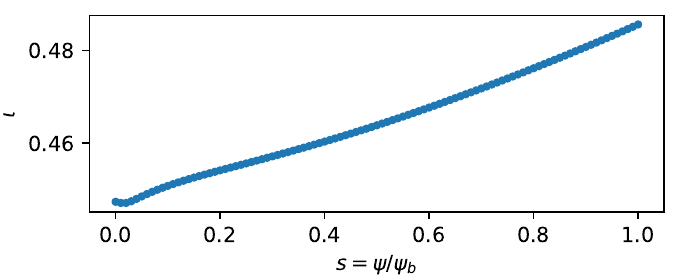}
        \end{subfigure} 
    \end{minipage}
    \caption{Quasi-isodynamic stellarator with 1 field period and 8 coils per half-field period.
    Top: standard stellarator optimization approach where stage 1 and stage 2 optimizations were performed sequentially (left) and the single-stage optimization result (right).
    Lower Left: Superposition of magnetic surfaces at constant cylindrical toroidal angle $\phi$ of the QFM and the final single-stage equilibrium, as well as the Poincaré plot resulting from tracing magnetic field lines in the obtained coils.
    Middle Right: Contours of constant magnetic field strength on a surface at $s=0.495$ in Boozer coordinates $(\theta, \varphi)$. Bottom Right: profile of rotational transform $\iota$.
    \label{fig:properties_QIncoils8nfp1}
    }
\end{figure}

We now present the optimization of a one-field period quasi-isodynamic stellarator with an aspect ratio of $A=7.0$. To achieve this goal, we used 8 independent coils with a total of $N_F=16$ Fourier modes, each coil with a maximum length of $L_{\text{max}}=5.5$, a maximum value of $\kappa_{\text{max}}=10.0$, and a minimum distance between coils of 0.12.
The results of our optimization are displayed in \cref{fig:properties_QIncoils8nfp1} (top), where we show the configuration obtained from stage 1 and stage 2 independent optimizations. The residuals of the quasi-isodynamic objective function $f_{\text{QI}}$ in stage 1 were $2.4 \times 10^{-3}$, while the residuals of the squared flux $f_{\text{QF}}$ in stage 2 were $2.9 \times 10^{-6}$.
In the single-stage optimization shown in \cref{fig:properties_QIncoils8nfp1} (top right), the residuals of the quasi-isodynamic objective function $f_{\text{QI}}$ remained mostly unchanged at $2.9 \times 10^{-3}$, while the squared flux improved to $4.4 \times 10^{-7}$.
We then run VMEC in fixed-boundary mode using the QFM surfaces obtained from the stage 1 and stage 2 independent optimizations and the single-stage approach.
Using the stage-2 QFM surface as a boundary, fixed-boundary VMEC failed to converge to a reasonable force residual.
However using the single-stage QFM surface as a boundary, VMEC converged and yields $f_{\text{QI}}=3.2\times10^{-3}$.
Thus, the traditional two-stage approach did not produce a usable result, whereas the single-stage method did.
The figures in \cref{fig:properties_QIncoils8nfp1} provide further insight into the effectiveness of our optimization. On the lower left, we see that the reduction in the squared flux results in an agreement between the single-stage fixed boundary equilibrium, a fixed boundary equilibrium based on the QFM surface, and the Poincaré plots. On the right, we display the contours of constant magnetic field at the $s=0.495$ surface in Boozer coordinates, which show the quasi-isodynamic character of this configuration, and its rotational transform profile.

\section{Conclusions}
\label{sec:conclusions}

The proposed single-stage optimization of both physics and engineering goals in coil systems provides a critical leap forward in the advancement of plasma physics and magnetic confinement. Our approach, which relies on fixed boundary equilibria, offers a significantly more efficient and versatile solution compared to previous numerical implementations based on free-boundary calculations. This is evident from the numerical examples presented in this paper, which demonstrate the effectiveness of the proposed method in obtaining various types of equilibria for stellarators with a reduced number of coils.

Our method directly balances plasma physics and coil engineering objectives by introducing a quadratic flux term in the objective function, resulting in consistency between the plasma shape and coil shapes.
We combine finite difference derivatives of the MHD equilibrium with analytic derivatives of the coils, reducing the number of finite difference steps required.
This approach is applicable to a wide range of vacuum and finite plasma pressure stellarator equilibria, including codes that do not yet have free boundary functionality, such as GVEC.
Furthermore, we require only one surface evaluation of the magnetic field from coils per optimization iteration, reducing computational time compared to volumetric evaluations used in other methods.
Compared to the free-boundary single-stage approach, our method is more efficient and adaptable.

We plan to extend our proposed method to investigate plasmas with finite plasma pressure.
This will be critical to understanding the behavior of plasma in magnetic confinement systems and will provide valuable information to design next-generation fusion reactors.
Fortunately, the only required modifications with respect to the present work are the inclusion of a target magnetic field in the quadratic flux term in the objective function, calculated using the virtual casing principle \cite{Shafranov1972} at each optimization step, and a single free-boundary MHD calculation at the end of the optimization (instead of the QFM surface approach used here) to assess the results.
{We also mention that, for the case of finite-beta plasmas, the profiles of pressure and current (or $\iota$) can be added as degrees of freedom, therefore extending the scope of the current method.}

{\it Data Availability}.---
The data that support the findings of this study are openly available in Zenodo at https://doi.org/10.5281/zenodo.7655077, reference number 7655077, and on GitHub at https://github.com/rogeriojorge/single\_stage\_optimization.

\section{Acknowledgments}
\label{sec:acknowledgments}

We thank the SIMSOPT team for their invaluable contributions.
R. J. is supported by the Portuguese FCT - Fundação para a Ciência e Tecnologia, under grant 2021.02213.CEECIND. 
The optimization studies were carried out using the
EUROfusion Marconi supercomputer facility.
This work has been carried out within the framework of the EUROfusion Consortium, funded by the European Union via the Euratom Research and Training Programme (Grant Agreement No 101052200 — EUROfusion). Views and opinions expressed are however those of the author(s) only and do not necessarily reflect those of the European Union or the European Commission. Neither the European Union nor the European Commission can be held responsible for them. IST activities also received financial support from FCT through projects UIDB/50010/2020 and UIDP/50010/2020.
This work was supported by a grant from the Simons Foundation (560651, ML).

\section*{References}
\bibliographystyle{unsrt}
\bibliography{library}

\end{document}